\documentclass[twocolumn,amsmath,amssymb]{revtex4}

\usepackage[english]{babel}
\usepackage{bm}
\usepackage{xcolor}
\usepackage{graphicx}
\usepackage{subfigure}
\usepackage{multirow}
\usepackage{rotating}

\usepackage{array}
\newcolumntype{H}{>{\setbox0=\hbox\bgroup}c<{\egroup}@{}}
\newcolumntype{G}{@{}>{\lrbox0}l<{\endlrbox}}

\definecolor{rouge}{RGB}{219,60,2} 
\definecolor{saumon}{RGB}{237,127,16} 
\definecolor{vert}{RGB}{130,180,0} 
\definecolor{brun}{RGB}{110,70,30} 

\newcommand{\mr}{\mathbf{r}}

\newcommand{\mx}{\mathbf{x}}
\newcommand{\mA}{\mathbf{A}}
\newcommand{\mB}{\mathbf{B}}

\newcommand{\mX}{\mathbf{X}}
\newcommand{\mY}{\mathbf{Y}}

\newcommand{\beq}{\begin{equation}}
\newcommand{\eeq}{\end{equation}}

\newcommand{\HF}{\text{HF}}

\newcommand{\IP}{\text{IP}}

\renewcommand{\d}{\text{d}}

\renewcommand{\H}{\text{H}}
\newcommand{\Hxc}{\text{Hxc}}
\newcommand{\Hx}{\text{Hx}}

\renewcommand{\b}[1]{\ensuremath{\mathbf{#1}}}

\newcommand{\mrm}{\mathrm}

\newcommand{\x}{\text{x}}
\renewcommand{\c}{\text{c}}
\newcommand{\ee}{\text{ee}}
\newcommand{\tc}{\text{c}}

\renewcommand{\H}{\ensuremath{\text{H}}}
\renewcommand{\b}[1]{\ensuremath{\mathbf{#1}}}

\newcommand{\lr}{\ensuremath{\text{lr}}}
\newcommand{\sr}{\ensuremath{\text{sr}}}

\newcommand{\pp}{\mrm{pp/hh}}
\newcommand{\ph}{\mrm{ph/hp}}

\long\def\ignore#1{}

\newcommand{\mv}{w_\text{ee}}
\newcommand{\barv}{\bar{w}_\text{ee}}

\DeclareMathOperator{\erf}{erf}

\begin{document}
\title{Range-separated time-dependent density-functional theory with a frequency-dependent second-order Bethe-Salpeter correlation kernel}
\author{Elisa Rebolini\footnote{Present address: Centre for Theoretical and Computational Chemistry, Department of Chemistry, University of Oslo, P.O. Box 1033 Blindern, N-0315 Oslo, Norway}
}\email{elisa.rebolini@kjemi.uio.no}
\author{Julien Toulouse}\email{julien.toulouse@upmc.fr}
\affiliation{
Sorbonne Universit\'es, UPMC Univ Paris 06, CNRS, Laboratoire de Chimie Th\'eorique, 4 place Jussieu, F-75005, Paris, France
}

\date{January 25, 2016}

\begin{abstract}
We present a range-separated linear-response time-dependent density-functional theory (TDDFT) which combines a density-functional approximation for the short-range response kernel and a frequency-dependent second-order Bethe-Salpeter approximation for the long-range response kernel. This approach goes beyond the adiabatic approximation usually used in linear-response TDDFT and aims at improving the accuracy of calculations of electronic excitation energies of molecular systems. A detailed derivation of the frequency-dependent second-order Bethe-Salpeter correlation kernel is given using many-body Green-function theory. Preliminary tests of this range-separated TDDFT method are presented for the calculation of excitation energies of the He and Be atoms and small molecules (H$_2$, N$_2$, CO$_2$, H$_2$CO, and C$_2$H$_4$). The results suggest that the addition of the long-range second-order Bethe-Salpeter correlation kernel overall slightly improves the excitation energies.
\end{abstract}

\maketitle

\section{Introduction}

Linear-response time-dependent density-functional theory (TDDFT)~\cite{GroKoh-PRL-85,Cas-INC-95} is nowadays one of the most popular approaches for calculating excitation energies and other response properties of electronic systems. Within the usual adiabatic semilocal density-functional approximations (DFAs), linear-response TDDFT usually provides reasonably accurate low-lying valence electronic excitation energies of molecular systems at a low computational cost. However, these usual adiabatic semilocal DFAs have serious failures. In particular, they give largely underestimated Rydberg~\cite{CasJamCasSal-JCP-98} and charge-transfer~\cite{DreWeiHea-JCP-03} excitation energies and they do not account for double (or multiple) excitations~\cite{MaiZhaCavBur-JCP-04}.

The problem with Rydberg and charge-transfer excitation energies is alleviated with the use of hybrid approximations in linear-response TDDFT~\cite{BauAhl-CPL-96}, which combine a Hartree-Fock (HF) exchange response kernel with a DFA exchange-correlation response kernel. This problem is essentially solved with range-separated hybrid approximations~\cite{TawTsuYanYanHir-JCP-04,YanTewHan-CPL-04,LivBae-PCCP-07,RebSavTou-MP-13}, introducing a long-range HF exchange kernel. Research in linear-response TDDFT now aims at an increasingly higher accuracy and reliability, and in particular the inclusion of the effects of the double excitations. Examples of recent developments are: the dressed TDDFT approach (combining TDDFT and the polarization-propagator approach)~\cite{Cas-JCP-05,HuiCas-ARX-10,CasHui-TCC-15}, double-hybrid TDDFT methods (combining TDDFT and configuration-interaction singles with doubles correction [CIS(D)])~\cite{GriNee-JCP-07}, and range-separated TDDFT approaches in which the long-range response is treated with density-matrix functional theory (DMFT)~\cite{Per-JCP-12}, multiconfiguration self-consistent-field (MCSCF) theory~\cite{FroKneJen-JCP-13}, or the second-order polarization-propagator approximation (SOPPA)~\cite{HedHeiKneFroJen-JCP-13}.

In condensed-matter physics, the Bethe-Salpeter equation (BSE) applied within the $GW$ approximation (see, e.g., Refs.~\onlinecite{Str-RNC-88,RohLou-PRB-00,OniReiRub-RMP-02}) is often considered as the most successful approach to overcome the limitations of TDDFT. Although it has been often used to describe excitons (bound electron-hole pair) in periodic systems, it is also increasingly applied to calculations of electronic excitation energies in finite molecular systems~\cite{Roh-IJQC-00,GroRohMitLouCoh-PRL-01,TiaChe-SSC-05,HahSchBec-PRB-05,TiaChe-PRB-06,TiaKenHooReb-JCP-08,GruMarGon-NL-09,MaRohMol-PRB-09,MaRohMol-JCTC-10,RocLuGal-JCP-10,GruMarGon-CMS-11,BlaAtt-APL-11,PalPavHubSch-EPJB-11,BauAndRoh-JCTC-12,DucDeuBla-PRL-12,FabDucDeuBla-PRB-12,RebTouSav-INC-13,DucBla-PRB-13,FabBouDucAttBla-JCP-13,BouJacDucBla-JCTC-14,KorBouDucBlaMarBot-JCTC-14,NogHiyAkiKog-JCP-14,BouChiLegDucBlaJac-JCTC-14,HirNogSug-PRB-15,RabBaeNeu-PRB-15,JacDucBla-JCTC-15,BruHamNea-JCP-15,LjuKovFerFoeSan-PRB-15,CocDra-PRB-15}. In particular, the BSE approach was shown to give accurate charge-transfer excitation energies in molecules~\cite{RocLuGal-JCP-10,BlaAtt-APL-11,BauAndRoh-JCTC-12,DucDeuBla-PRL-12,FabDucDeuBla-PRB-12,DucBla-PRB-13,FabBouDucAttBla-JCP-13}, and when used with a frequency-dependent kernel it is in principle capable of describing double excitations~\cite{RomSanBerSotMolReiOni-JCP-09,SanRomOniMar-JCP-11,PalPavHubSch-EPJB-11}. The drawbacks of the standard BSE approach with the usual approximations are the need to first perform a computationally demanding $GW$ quasiparticle calculation, and an observed loss of accuracy for small molecules~\cite{HirNogSug-PRB-15} which is probably due to self screening.

In this work, we explore the combination of TDDFT and BSE approaches based on a range separation of the electron-electron interaction. More specifically, we propose a range-separated TDDFT approach in which the long-range response is treated with a frequency-dependent second-order Bethe-Salpeter-equation (BSE2) correlation kernel. The BSE2 approximation was recently introduced by Zhang \emph{et al.}~\cite{ZhaSteYan-JCP-13} within the Tamm-Dancoff approximation (TDA)~\cite{HirHea-CPL-99}. Compared to the standard BSE approach with the $GW$ approximation, the BSE2 approximation keeps only second-order terms with respect to the electron-electron interaction, including second-order exchange terms which makes it free from self screening. It is an appropriate approximation for finite molecular systems with relatively large gaps. Building on the work of Sangalli {\it et al.}~\cite{SanRomOniMar-JCP-11}, we provide an alternative and more general derivation of the BSE2 approximation and we apply it to the range-separated case. We present preliminary tests of this range-separated TDDFT method for the calculation of excitation energies of the He and Be atoms and some small molecules (H$_2$, N$_2$, CO$_2$, H$_2$CO, and C$_2$H$_4$).

In this range-separated TDDFT approach, an adiabatic semilocal DFA is used only for the short-range part of the exchange-correlation kernel, while a frequency dependence is introduced in the long-range part of the correlation kernel. This is motivated by the fact that the exact exchange kernel becomes spatially local and frequency independent in the limit of a very short-range interaction~\cite{RebSavTou-MP-13,FraLupTou-JJJ-XX}, so that the adiabatic local-density approximation (LDA) becomes exact in this limit. Similarly, the short-range part of the exact correlation kernel is expected to be more spatially local and less frequency dependent than its long-range counterpart, so that an adiabatic semilocal DFA is expected to be accurate when restricted to the short-range part of the correlation kernel, as it happens for the ground-state correlation density functional~\cite{TouColSav-PRA-04}. 

Similarly to the ground-state case where second-order perturbation theory is appropriate for describing the long-range part of the correlation energy of systems with large enough gaps~\cite{AngGerSavTou-PRA-05}, the BSE2 approximation is expected to be appropriate for describing the long-range part of the response of such systems. Moreover, in comparison to the original full-range BSE2 scheme, the restriction of the BSE2 approximation to the long-range part leads to potential practical and computational advantages: (1) eliminating the need to do a first $GW$ quasiparticle calculation since range-separated hybrid approximations provide orbital energies that are already close to quasiparticle energies~\cite{TsuSonSuzHir-JCP-10,KroSteRefBae-JCTC-12}; and (2) speeding up the computation of the BSE2 correlation kernel by using multipole expansions for the long-range two-electron integrals~\cite{HetSchStoWer-JCP-00}.

This paper is organized as follows. In Section~\ref{sec:rsTDDFT}, we summarize the main equations of linear-response TDDFT with range separation. In Section~\ref{sec:BSE2}, we provide a full derivation of the frequency-dependent BSE2 correlation kernel without using the TDA, giving expressions in terms of space-spin coordinates and in a spin-orbital basis. Section~\ref{sec:RSH_BSE} explains how we practically perform the calculations and gives computational details for the systems tested. The results are given and discussed in Section~\ref{sec:results dyn}. Finally, Section~\ref{sec:conclu} contains our conclusions.

\section{Range-separated time-dependent density-functional theory}
\label{sec:rsTDDFT}

As a relatively straightforward extension of linear-response TDDFT~\cite{GroKoh-PRL-85}, in range-separated TDDFT~\cite{FroKneJen-JCP-13,RebSavTou-MP-13}, the inverse of the frequency-dependent linear-response function is expressed as
\begin{eqnarray}
\label{eq:chim1-RSTDDFT}
\chi^{-1}(\b{x}_1,\b{x}_2;\b{x}_1',\b{x}_2';\omega) &=& (\chi^{\lr})^{-1}(\b{x}_1,\b{x}_2;\b{x}_1',\b{x}_2';\omega) 
\nonumber\\
&&- f_{\Hxc}^{\sr}(\b{x}_1,\b{x}_2;\b{x}_1',\b{x}_2';\omega),
\end{eqnarray}
where $\b{x}=(\b{r},\sigma)$ stands for space-spin coordinates.
In this expression, $\chi^{\lr}(\b{x}_1,\b{x}_2;\b{x}_1',\b{x}_2';\omega)$ is the linear-response function associated with the long-range (lr) interacting Hamiltonian
\begin{eqnarray}
\hat{H}^{\lr} = \hat{T} + \hat{V}_{\text{ne}} + \hat{W}_{\ee}^{\lr} + \hat{V}_{\Hxc}^{\sr},
\label{Hlr}
\end{eqnarray}
where $\hat{T}$ is the kinetic-energy operator, $\hat{V}_{\text{ne}}$ is the nuclei-electron interaction operator, $\hat{W}_{\ee}^{\lr}$ is a long-range electron-electron interaction operator, and $\hat{V}_{\Hxc}^{\sr}$ is a corresponding short-range (sr) Hartree--exchange--correlation (Hxc) potential operator. Additionally, $f_{\Hxc}^{\sr}(\b{x}_1,\b{x}_2;\b{x}_1',\b{x}_2';\omega) = f_{\Hxc}^{\sr}(\b{x}_1,\b{x}_2;\omega) \delta (\b{x}_1,\b{x}_1') \delta (\b{x}_2,\b{x}_2')$ is the short-range Hxc kernel related to the functional derivative of the short-range Hxc potential with respect to the density (and $\delta$ is the delta function). In practice, the long-range electron-electron interaction is defined with the error function as $w_{\ee}^{\lr}(\b{r}_1,\b{r}_2)=\erf(\mu |\b{r}_1 -\b{r}_2|)/|\b{r}_1 -\b{r}_2|$, where the parameter $\mu$ controls the range of the interaction. Even though Eq.~(\ref{eq:chim1-RSTDDFT}) is written with functions depending on four space-spin coordinates for generality, range-separated TDDFT only gives exactly the diagonal part of the linear-response function $\chi(\b{x}_1,\b{x}_2;\omega) = \chi(\b{x}_1,\b{x}_2;\b{x}_1,\b{x}_2;\omega)$, just as in usual TDDFT.

In the time-dependent range-separated hybrid (TDRSH) scheme~\cite{RebSavTou-MP-13}, the long-range linear-response function $\chi^{\lr}(\omega)$ is calculated at the HF level. More precisely, the inverse of the long-range linear-response function is approximated as
\begin{eqnarray}
(\chi^{\lr})^{-1}(\b{x}_1,\b{x}_2;\b{x}_1^\prime,\b{x}_2^\prime;\omega) &\approx& (\chi_{0})^{-1}(\b{x}_1,\b{x}_2;\b{x}_1^\prime,\b{x}_2^\prime;\omega) 
\nonumber\\
& -& f_{\Hx,\HF}^{\lr}(\b{x}_1,\b{x}_2;\b{x}_1^\prime,\b{x}_2^\prime),
\label{chilrmuinvHF}
\end{eqnarray}
where $\chi_{0}(\omega)$ is the non-interacting linear-response function associated with the range-separated-hybrid (RSH) reference Hamiltonian~\cite{AngGerSavTou-PRA-05}
\begin{eqnarray}
\hat{H}_0 = \hat{T} +  \hat{V}_{\text{ne}} + \hat{V}_{\Hx,\HF}^{\lr} + \hat{V}_{\Hxc}^{\sr},
\end{eqnarray}
with the long-range HF potential operator $\hat{V}_{\Hx,\HF}^{\lr}$, and $f_\Hx^{\lr}(\b{x}_1,\b{x}_2;\b{x}_1^\prime,\b{x}_2^\prime)$ is the corresponding long-range HF kernel. The latter is the sum of a long-range Hartree kernel,
\begin{eqnarray}
f_\H^{\lr}(\b{x}_1,\b{x}_2;\b{x}_1^\prime,\b{x}_2^\prime) = w^{\lr}_{\ee}(\b{r}_1,\b{r}_2) \delta(\b{x}_1,\b{x}_1^\prime) \delta(\b{x}_2,\b{x}_2^\prime),
\end{eqnarray}
and a long-range HF exchange kernel,
\begin{eqnarray}
f_{\x,\HF}^{\lr}(\b{x}_1,\b{x}_2;\b{x}_1^\prime,\b{x}_2^\prime) = -w^{\lr}_{\ee}(\b{r}_1,\b{r}_2) \delta(\b{x}_1,\b{x}_2^\prime) 
\delta(\b{x}_2,\b{x}_1^\prime).\;
\end{eqnarray}

To go beyond the HF level, it was proposed to calculate $\chi^{\lr}(\omega)$ at the linear-response MCSCF level~\cite{FroKneJen-JCP-13} or at the SOPPA level~\cite{HedHeiKneFroJen-JCP-13}. In the present work, we explore the recently proposed BSE2 approximation~\cite{ZhaSteYan-JCP-13}. We thus propose to approximate the inverse of the long-range linear-response function as
\begin{eqnarray}
\label{eq:chim1-RSH-BSE2}
(\chi^{\lr})^{-1}(\b{x}_1,\b{x}_2;\b{x}_1^\prime,\b{x}_2^\prime;\omega) \approx (\chi_{0})^{-1}(\b{x}_1,\b{x}_2;\b{x}_1^\prime,\b{x}_2^\prime;\omega) 
\nonumber\\
- f_{\Hx,\HF}^{\lr}(\b{x}_1,\b{x}_2;\b{x}_1^\prime,\b{x}_2^\prime) - f_{\c,\text{BSE2}}^{\lr}(\b{x}_1,\b{x}_2;\b{x}_1^\prime,\b{x}_2^\prime;\omega),
\end{eqnarray}
with the long-range BSE2 frequency-dependent correlation kernel $f_{\c,\text{BSE2}}^{\lr}(\omega)$ for which we offer an alternative and more general derivation compared to Ref.~\onlinecite{ZhaSteYan-JCP-13}.

\section{Second-order Bethe-Salpeter correlation kernel}
\label{sec:BSE2}

In this Section, we provide a derivation of the BSE2 correlation kernel. For more details, see Ref.~\onlinecite{Reb-THESIS-14_eng}. We consider an arbitrary electron-electron interaction $w_\ee$ in the derivation instead of the long-range one.

\subsection{Second-order correlation self-energy}

\begin{figure}
\includegraphics[scale=1]{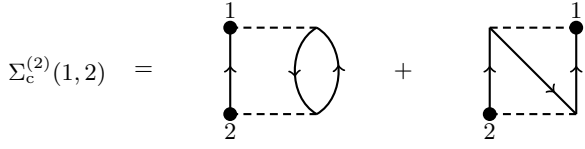}
 \caption{Feynman diagrams of the second-order correlation self-energy $\Sigma_\c^{(2)}(1,2)$. The time axis is vertical. The dots represent the outer variables $1$ and $2$. Horizontal dashed lines represent electron-electron interactions $\mv$. Arrowed lines represent one-particle Green functions $G$. The first diagram is the direct contribution of Eq.~(\ref{eq:sigma_c_d def}) and the second diagram is the exchange contribution of Eq.~(\ref{eq:sigma_c_x def}).
}
\label{fig:sigma_tikz}
\end{figure}

The starting point is the second-order correlation self-energy as a functional of the one-electron Green function $G(1,2)$ where $1=(\b{x}_1,t_1)$ and $2=(\b{x}_2,t_2)$ stand for space-spin-time coordinates (see, e.g., Ref.~\onlinecite{RebTouSav-INC-13})
\begin{equation}
\begin{split}
  \label{eq:sigma_c def}
  \Sigma_\tc^{(2)}&(1,2) = i \int \d3\d3' \d4 \d4'\d5 \d5' \; G(3,3') \\
  &\barv(3',4;2,4') \chi_{\IP}(4',5;4,5')
  \mv(5',1;5,3), \\
\end{split}
\end{equation}
where $\mv(1,2;1',2')=\mv(1,2) \delta(1,1')\delta(2,2')$ is an arbitrary electron-electron interaction, $\barv(1,2;1',2') = \mv(1,2;1',2') -\mv(2,1;1',2')$ is the corresponding antisymmetrized interaction, and $\chi_\IP(1,2;1',2') = -iG(1,2')G(2,1')$ is the independent-particle (IP) four-point linear-response function. The presence of the antisymmetrized interaction $\barv$ in Eq.~(\ref{eq:sigma_c def}) means that the second-order correlation self-energy can be decomposed as $\Sigma_\c^{(2)} = \Sigma_\tc^{(2\text{d})} + \Sigma_\tc^{(2\x)}$ with a direct contribution
\begin{eqnarray}
\label{eq:sigma_c_d def}
\Sigma_\tc^{(2\text{d})}(1,2) &=& i \; G(1,2) \, \int \d 3 \d 4 \; \mv(2,3)
\nonumber\\
&&\times \chi_{\IP}(3,4;3,4) \mv(4,1), 
\end{eqnarray}
and an exchange contribution
\begin{eqnarray}
   \label{eq:sigma_c_x def}
 \Sigma_\tc^{(2\x)}(1,2) &=&- i \int \d3 \d4 \; G(1,3) \mv(2,3) 
\nonumber\\
&&\times \chi_{\IP}(3,4;2,4) \mv(4,1).
\end{eqnarray}
The Feynman diagrams of these terms are represented in Figure~\ref{fig:sigma_tikz}.

\subsection{Second-order Bethe-Salpeter correlation kernel in the time domain}

\begin{figure*}
\includegraphics[scale=1]{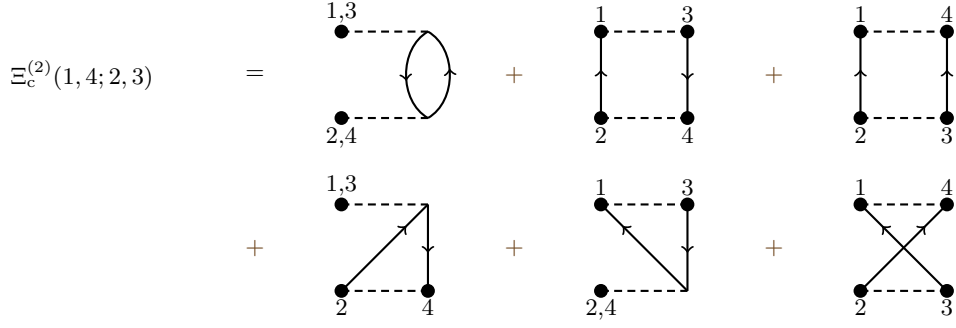}
  \caption{Feynman diagrams of the second-order Bethe-Salpeter correlation kernel $\Xi_{\c}^{(2)}(1,4;2,3)$. The three upper diagrams are the direct contributions of Eq.~(\ref{eq:Xic2d}) and the three lower diagrams are the exchange contributions of Eq.~(\ref{eq:Xic2x}). The four diagrams on the left correspond to ph/hp terms and the two diagrams on the right correspond to pp/hh diagrams. Since the kernel is the functional derivative of the self-energy with respect to the Green function, these diagrams can be obtained from the ones of Figure~\ref{fig:sigma_tikz} by removing one of the arrowed lines.
    \label{fig:kernel_derivation_tikz}}
\end{figure*}

The second-order Bethe-Salpeter correlation kernel is defined as the functional derivative of the second-order correlation self-energy with respect to the Green function
\begin{equation}
  \Xi_{\c}^{(2)}(1,4;2,3) =
    i\dfrac{\delta \Sigma_{\c}^{(2)}(1,2)}{\delta G(3,4)}.
\end{equation}
Taking the derivative of Eq.~(\ref{eq:sigma_c def}) generates three terms
\begin{eqnarray}
\Xi_{\c}^{(2)}(1,4;2,3) = \phantom{xxxxxxxxxxxxxxxxxxxxxxxxxxxxx}
\nonumber\\
- \int \d 5 \d 5' \d 6 \d 6' \barv(4,5;2,5') \chi_\text{IP}(5',6;5,6') \mv(6',1;6,3)
\nonumber\\
- \int \d 5 \d 5' \d 6 \d 6' \barv(5,4;2,6)  \chi_\text{IP}(5',6;6',5) \mv(6',1;3,5')
\nonumber\\
- \int \d 5 \d 5' \d 6 \d 6' \barv(5,6;2,3) \chi_\text{IP}(6',5';6,5) \mv(4,1;5',6'),
\nonumber\\
\end{eqnarray}
which, as done for the correlation self-energy, can be decomposed as $\Xi_{\c}^{(2)} = \Xi_{\c}^{(2\text{d})} + \Xi_{\c}^{(2\text{x})}$, with a direct contribution
\begin{eqnarray}
\lefteqn{\Xi_{\c}^{(2\text{d})}(1,4;2,3) =}&
\nonumber\\
&&\!- \, \delta(1,3) \delta(2,4) \! \int \d 5 \d 6 \, \mv(2,5) 
\chi_\text{IP}(5,6;5,6) \mv(6,1)
\nonumber\\
&&- \, \mv(2,4)  \chi_\text{IP}(1,4;3,2) \mv(3,1)
\nonumber\\
&&- \, \mv(2,3) \chi_\text{IP}(1,4;3,2) \mv(4,1),
\label{eq:Xic2d}
\end{eqnarray}
and an exchange contribution
\begin{eqnarray}
\lefteqn{\Xi_{\c}^{(2\text{x})}(1,4;2,3) =}&&
\nonumber\\
&&\delta(1,3) \int \d 5 \mv(2,4) \chi_\text{IP}(4,5;2,5) \mv(5,1)
\nonumber\\
&+& \delta(2,4) \int \d 5 \mv(2,5)  \chi_\text{IP}(1,5;3,5) \mv(3,1)
\nonumber\\
&+& \mv(2,3) \chi_\text{IP}(1,4;2,3) \mv(4,1).
\label{eq:Xic2x}
\end{eqnarray}
The Feynman diagrams of these six terms are represented in Figure~\ref{fig:kernel_derivation_tikz}. Similar kernel diagrams are shown in Ref.~\onlinecite{DahLee-PRL-07}.

Introducing explicitly the time variables, using an instantaneous spin-independent electron-electron interaction $\mv(1,2)=\mv(\mr_1,\mr_2) \delta(t_1,t_2)$ and time-translation invariance, we found that the second-order Bethe-Salpeter correlation kernel is composed of a particle-hole/hole-particle (ph/hp) part and a particle-particle/hole-hole (pp/hh) part, which non-trivially depend on only one time difference $t_1-t_2$,
\begin{equation}
  \begin{split}
\Xi&_{\c}^{(2)}(\mx_1 t_1,\mx_4 t_4;\mx_2 t_2,\mx_3 t_3) \\
 = & \, \delta(t_1,t_3) \delta(t_2,t_4) \Xi_{\c}^{(2,\ph)}(\mx_1,\mx_4;\mx_2,\mx_3; t_1-t_2)\\
   & +  \delta(t_1,t_4) \delta(t_2,t_3) \Xi_{\c}^{(2,\pp)}(\mx_1,\mx_4;\mx_2,\mx_3; t_1-t_2),
  \end{split}
\label{eq:Xic2times}
\end{equation}
with the ph/hp kernel
\begin{equation}
\begin{split}
\Xi&_{\c}^{(2,\ph)}(\mx_1,\mx_4;\mx_2,\mx_3; \tau) =\\
 = & \, - \delta(\mx_1,\mx_3) \delta(\mx_2,\mx_4) \! \int \d \mx_5 \d \mx_6 \, \mv(\mr_2,\mr_5) \\
   & \times \chi_\text{IP}(\mx_5,\mx_6;\mx_5,\mx_6;-\tau) \mv(\mr_6,\mr_1)\\
   & - \mv(\mr_2,\mr_4) \chi_\text{IP}(\mx_1,\mx_4;\mx_3, \mx_2;\tau) \mv(\mr_3,\mr_1)\\
   & + \delta(\mx_1,\x_3) \int \d \mx_5 \mv(\mr_2,\mr_4) \\
   & \times \chi_\text{IP}(\mx_4,\mx_5;\mx_2,\mx_5;-\tau) \mv(\mr_5,\mr_1) \\
   & + \delta(\mx_2,\mx_4) \int \d \mx_5 \mv(\mr_2,\mr_5)\\
   & \times \chi_\text{IP}(\mx_1,\mx_5;\mx_3,\mx_5;\tau) \mv(\mr_3,\mr_1),
\end{split}
\label{eq:Xic2ph}
\end{equation}
and the pp/hh kernel
\begin{equation}
\begin{split}
\Xi&_{\c}^{(2\text{pp/hh})}(\mx_1,\mx_4;\mx_2,\mx_3; \tau) \\
 =  & -  \mv(\mr_2,\mr_3) \chi_\text{IP}^\pp(\mx_1,\mx_4;\mx_3,\mx_2;\tau) \mv(\mr_4,\mr_1)\\
    & + \mv(\mr_2,\mr_3) \chi_\text{IP}^\pp(\mx_1,\mx_4;\mx_2,\mx_3;\tau) \mv(\mr_4,\mr_1),
\end{split}
\label{eq:Xic2pp}
\end{equation}
where $\chi_\text{IP}(\mx_1,\mx_2;\mx_1',\mx_2'; \tau=t_1-t_2) = \chi_\text{IP}(\mx_1 t_1,\mx_2 t_2;\mx_1' t_1,\mx_2' t_2)$ is the IP (ph/hp) linear-response function and $\chi_\text{IP}^{\pp}(\mx_1,\mx_2;\mx_1',\mx_2'; \tau = t_1-t_2) = \chi_\text{IP}(\mx_1 t_1,\mx_2 t_1;\mx_1' t_2,\mx_2' t_2)$ is the IP pp/hh linear-response function. As the names suggest, the former describes the independent propagation of one particle and one hole, and the latter describes the independent propagation of either two particles or two holes, depending on the sign of $t_1-t_2$. Because of the different delta functions on the time variables in Eq.~(\ref{eq:Xic2times}), the ph/hp and pp/hh contributions need to be treated separately.

\subsection{Effective second-order Bethe-Salpeter correlation kernel in the frequency domain}

The Bethe-Salpeter kernel that we derived must be used in the general Bethe-Salpeter equation in the time domain which is~\cite{BayKad-PR-61,Bay-PR-62}
\label{sec:dyn_kernel}
\begin{equation}
  \begin{split}
    \chi(1,2;1',2') = & \,\chi_{\IP}(1,2;1',2') + \int \d 3 \d 4 \d 5 \d 6 \\
    & \chi_{\IP}(1,4;1',3)  \Xi_{\Hxc}(3,6;4,5) \chi(5,2;6,2'),
  \end{split}
\end{equation}
where $\chi(1,2;1',2')$ is the interacting four-point linear-response function and $\Xi_{\Hxc}$ is the Bethe-Salpeter Hxc kernel. Written explicitly with time variables, and setting $t_1'=t_1^+$ and $t_2'=t_2^+$ (where $t^+ = t + 0^+$ refers to a time variable with an infinitesimal positive shift) to extract the (ph/hp) linear-response function, the equation becomes
\begin{equation}
  \begin{split}
    &\chi(\mx_1t_1,\mx_2t_2; \mx_1't_1^+,\mx_2't_2^+) 
    =  \chi_{\IP}(\mx_1t_1,\mx_2t_2;\mx_1't_1^+,\mx_2't_2^+) \\
    & + \!\! \int \!\! \d\mx_3\d t_3 \d\mx_4 \d t_4 \d\mx_5 \d t_5 \d\mx_6 \d t_6
    \chi_{\IP}(\mx_1t_1,\mx_4t_4;\mx_1't_1^+,\mx_3t_3) \\
    & \qquad \Xi_{\Hxc}(\mx_3t_3,\mx_6t_6;\mx_4t_4,\mx_5t_5) 
    \chi(\mx_5t_5,\mx_2t_2;\mx_6t_6,\mx_2't_2^+),
  \end{split}
\label{eq:BSEtimes}
\end{equation}
where $\Xi_{\Hxc}(\mx_3t_3,\mx_6t_6;\mx_4t_4,\mx_5t_5) = f_{\Hx,\HF}(\mx_3,\mx_6;\mx_4,\mx_5) + \Xi_{\c}^{(2)}(\mx_3t_3,\mx_6t_6;\mx_4t_4,\mx_5t_5)$ is taken as the sum of time-independent HF kernel $f_{\Hx,\HF}$ and the second-order correlation kernel $\Xi_{\c}^{(2)}$. Because of the time dependence in $\Xi_{\c}^{(2)}$, in Eq.~(\ref{eq:BSEtimes}) the time variables $t_3$ and $t_4$ cannot be equated, and neither can be the time variables $t_5$ and $t_6$. Consequently, Eq.~(\ref{eq:BSEtimes}) is not a closed equation for the (ph/hp) linear-response function $\chi(\mx_1t_1,\mx_2t_2; \mx_1't_1^+,\mx_2't_2^+)$. 

To close the equation, Zhang {\it et al.}~\cite{ZhaSteYan-JCP-13} followed Strinati~\cite{Str-RNC-88} and used an explicit time-dependent form in the TDA for the ph/hp amplitudes~\cite{CsaTayYar-INC-71,Str-RNC-88} with which $\chi$ in Eq.~(\ref{eq:BSEtimes}) can be expressed. Here, instead, following Sangalli {\it et al.}~\cite{SanRomOniMar-JCP-11} (see also Ref.~\onlinecite{RomSanBerSotMolReiOni-JCP-09}), we work in Fourier space and define an effective kernel depending on only one frequency without using the TDA. Introducing the decomposition of $\Xi_{\c}^{(2)}$ in ph/hp and pp/hh terms given in Eq.~(\ref{eq:Xic2times}) and Fourier transforming Eq.~(\ref{eq:BSEtimes}) gives
\begin{equation}
  \begin{split}
    &\chi(\omega) = \chi_{\IP}(\omega) + \chi_{\IP}\left(\omega
    \right) \, f_{\Hx} \, \chi(\omega) \\
    & + \int \dfrac{\d\omega'}{2\pi} \dfrac{\d\omega''}{2\pi}
    \chi_{\IP}\left(\omega',\omega \right)  
    \Xi_{\tc}^{(2,\ph)}(\omega' - \omega'') \chi(
    \omega'',\omega) \\
    & + \int \dfrac{\d\omega'}{2\pi} \dfrac{\d\omega''}{2\pi} \chi_{\IP}(
    \omega',\omega) 
    \Xi_{\tc}^{(2,\pp)}(\omega'+\omega'') \chi(
    \omega'',\omega),
    \label{eq:BSE fourier transform}
  \end{split}
\end{equation}
where the space-spin variables have been dropped for conciseness (all the quantities depend on four space-spin variables), and the integrations over $\omega'$ and $\omega''$ are from $-\infty$ to $+\infty$. In this expression, $\chi(\omega',\omega)$ is the double Fourier transform
\begin{equation}
\chi(\omega',\omega) = \int \d\tau_1 \d\tau \, e^{i\omega' \tau_1} e^{i\omega \tau} \chi(\tau_1,\tau_2=0^-,\tau),
\end{equation}
where $\chi(\tau_1,\tau_2,\tau)$ corresponds to the response function $\chi(\mx_1t_1,\mx_2t_2; \mx_1't_1',\mx_2't_2')$ expressed with the time variables $\tau_1=t_1-t_1'$, $\tau_2=t_2-t_2'$, and $\tau=(t_1 +t_1')/2-(t_2 +t_2')/2$, and similarly for $\chi_\IP(\omega',\omega)$. As a special case, $\chi(\omega)$ is just the Fourier transform of the (ph/hp) linear-response function $\chi(\tau_1=0,\tau_2=0,\tau)$, and similarly for $\chi_\IP(\omega)$. Obviously, $\Xi_{\tc}^{(2,\ph)}(\omega)$ and $\Xi_{\tc}^{(2,\pp)}(\omega)$ are the Fourier transforms of $\Xi_{\tc}^{(2,\ph)}(\tau)$ and $\Xi_{\tc}^{(2,\pp)}(\tau)$ given in Eqs.~(\ref{eq:Xic2ph}) and~(\ref{eq:Xic2pp}), respectively. Eq.~(\ref{eq:BSE fourier transform}) can be rewritten as an effective Bethe-Salpeter equation involving only one frequency~\cite{SanRomOniMar-JCP-11}
\begin{equation} 
  \begin{split}
    \chi(\omega) 
    =& \, \chi_\IP(\omega) + \chi_\IP(\omega) \, f_{\Hx} \, \chi(\omega) + 
    \chi_\IP(\omega) \tilde{\Xi}_\tc^{(2)}(\omega) \chi(\omega),
  \end{split}
\label{eq:effectiveBSE}
\end{equation}
or, equivalently,
\begin{equation} 
\chi^{-1}(\omega) = \chi_\IP^{-1}(\omega) - f_{\Hx} - \tilde{\Xi}_\tc^{(2)}(\omega),
\label{eq:effectiveBSEinv}
\end{equation}
with an effective correlation kernel defined as
\begin{equation}
  \begin{split}
    &\tilde{\Xi}_\tc^{(2)}(\omega) = \chi_{\IP}^{-1}(\omega) \int \dfrac{\d\omega'}{2\pi}
    \dfrac{\d\omega''}{2\pi} \chi_{\IP}(\omega', \omega)\\ & \qquad
    \Xi_{\tc}^{(2,\ph)}(\omega' - \omega'')
    \chi(\omega'',\omega)\chi^{-1}(\omega) \\
    & + \chi_\IP^{-1}(\omega) \int \dfrac{\d\omega'}{2\pi}
    \dfrac{\d\omega''}{2\pi} \chi_{\IP}(\omega', \omega)\\ & \qquad
    \Xi_{\tc}^{(2,\pp)}(\omega' + \omega'')
    \chi(\omega'',\omega)\chi^{-1}(\omega).
\end{split}
\label{eq:tildeXic}
\end{equation}
To keep only second-order terms in Eq.~(\ref{eq:tildeXic}) we must replace both the IP linear-response function $\chi_\IP$ and interacting linear-response function $\chi$ by the non-interacting linear-response function $\chi_0$, and we finally arrive at the BSE2 correlation kernel
\begin{equation}
  \begin{split}
    &f_{\c,\text{BSE2}}(\omega) = \chi_{0}^{-1}(\omega) \int \dfrac{\d\omega'}{2\pi}
    \dfrac{\d\omega''}{2\pi} \chi_{0}(\omega', \omega)\\ & \qquad
    \Xi_{\tc}^{(2,\ph)}(\omega' - \omega'')
    \chi_0(\omega'',\omega)\chi_0^{-1}(\omega) \\
    & + \chi_0^{-1}(\omega) \int \dfrac{\d\omega'}{2\pi}
    \dfrac{\d\omega''}{2\pi} \chi_{0}(\omega', \omega)\\ & \qquad
    \Xi_{\tc}^{(2,\pp)}(\omega' + \omega'')
    \chi_0(\omega'',\omega)\chi_0^{-1}(\omega),
\end{split}
\label{eq:fcBSE2}
\end{equation}
where $\Xi_{\tc}^{(2,\ph)}$ and $\Xi_{\tc}^{(2,\pp)}$ are obtained from Eqs.~(\ref{eq:Xic2ph}) and~(\ref{eq:Xic2pp}) with the replacement of $\chi_\IP$ by $\chi_0$ as well.

We note that, in Eq.~(\ref{eq:effectiveBSEinv}), $\chi_\IP^{-1}(\omega)$ could also be expanded to second order, leading to self-energy (or quasiparticle) contributions to the effective kernel~\cite{SanRomOniMar-JCP-11}. However, in this work, we do not consider such self-energy contributions to the kernel.

\subsection{Expressions in a spin-orbital basis}

We now give expressions in the orthonormal canonical spin-orbital basis $\{\varphi_p\}$ of the reference non-interacting Hamiltonian. Any function $F(\mx_1,\mx_2;\mx_1',\mx_2')$ depending on four space-spin coordinates can be expanded in the basis of products of two spin orbitals, and its matrix elements are defined as
\begin{equation}
  \begin{split}
    F_{pq,rs} =& \,\int \d\mx_1 \d\mx_1' \d\mx_2 \d\mx_2'
    \varphi_p(\mx_1') \varphi_q^*(\mx_1) \\ &
    F(\mx_1,\mx_2;\mx_1',\mx_2') \varphi_r^*(\mx_2)
    \varphi_s(\mx_2'),
  \label{eq:convention xi}
  \end{split}
\end{equation}
where $p,q,r,s$ refer to any (occupied or virtual) spin orbital. In the following, the indices $i,j,k,l$ will refer to occupied spin orbitals and the indices $a,b,c,d$ to virtual spin orbitals.

Using the expression of the Fourier transform of the non-interacting (ph/ph) linear-response function,
\begin{equation}
\begin{split}
\chi_0&(\mx_1,\mx_2;\mx_1',\mx_2';\omega) = \\
&\sum_{kc} 
\dfrac{\varphi_k^*(\mx_1')\varphi_c(\mx_1)\varphi_c^*(\mx_2')\varphi_k(\mx_2) }
{\omega -(\varepsilon_c - \varepsilon_k) +i0^+} \\
-&\sum_{kc} \dfrac{\varphi_k^*(\mx_2')\varphi_c(\mx_2)\varphi_c^*(\mx_1')\varphi_k(\mx_1) }
{\omega +(\varepsilon_c - \varepsilon_k) -i0^+},
  \end{split}
\end{equation}
and of the non-interacting pp/hh linear-response function, 
\begin{equation}
 \begin{split}
\chi_0^\pp&(\mx_1,\mx_2;\mx_1',\mx_2';\omega) = \\
&\sum_{kl} 
\dfrac{\varphi_k^*(\mx_1')\varphi_l(\mx_1)\varphi_l^*(\mx_2')\varphi_k(\mx_2) }
{\omega -(\varepsilon_k + \varepsilon_l) -i0^+} \\
-& \sum_{cd} \dfrac{\varphi_c^*(\mx_1')\varphi_d(\mx_1)\varphi_d^*(\mx_2')\varphi_c(\mx_2) }
{\omega +(\varepsilon_c + \varepsilon_d) +i0^+},
  \end{split}
\end{equation}
where $\varepsilon_p$ are the spin-orbital energies, we find the matrix elements of the Fourier transform of the ph/hp second-order correlation kernel,
\begin{equation}
  \begin{split}
    \Xi_{\tc,pq,rs}^{(2,\ph)}(\omega) = & - \sum_{kc} \dfrac{ \langle
      rc||pk \rangle \langle kq||cs \rangle } {\omega - (\varepsilon_c
      - \varepsilon_k) + i0^+} \\
    &+ \sum_{kc} \dfrac{ \langle rk||pc
      \rangle \langle cq||ks \rangle} {\omega + (\varepsilon_c -
      \varepsilon_k) - i0^+},
    \label{eq:second-order correlation Xi}
\end{split}
\end{equation}
and of the pp/hh second-order correlation kernel,
\begin{equation}
  \begin{split}
    \Xi_{\tc,pq,rs}^{(2,\pp)}(\omega) =  
    -\dfrac{1}{2}\sum_{kl}\dfrac{ \langle qr||kl\rangle \langle lk||sp
      \rangle} {\omega - (\varepsilon_k+\varepsilon_l)-i0^+} \\
    +\dfrac{1}{2}\sum_{cd}\dfrac{\langle qr||cd\rangle \langle dc||sp
      \rangle } {\omega - (\varepsilon_c+\varepsilon_d) + i0^+}.
    \label{eq:second-order correlation Xi pp}
\end{split}
\end{equation}
where $\langle pq||rs\rangle = \langle pq|rs\rangle - \langle pq|sr\rangle$ are the antisymmetrized two-electron integrals associated with the interaction $\mv$.

The matrices of $\chi_{0}(\omega)$ and $\chi_{0}(\omega',\omega)$ are both diagonal with elements in the occupied-virtual/occupied-virtual spin-orbital product block given by
\begin{eqnarray}
\chi_{0,ia,ia}(\omega) =  \dfrac{1}{\omega - (\varepsilon_a - \varepsilon_i) + i0^+},
\end{eqnarray}
and (see Appendix~\ref{app})
\begin{eqnarray}
\chi_{0,ia,ia}(\omega',\omega) = i \; e^{i\omega' 0^+} \; \chi_{0,ia,ia}(\omega) \; \phantom{xxxxxxxxxxx}
\nonumber\\
\times
\left( \dfrac{1}{\omega'+\omega/2 - \varepsilon_a + i0^+} - \dfrac{1}{\omega'-\omega/2 - \varepsilon_i - i0^+}\right),
\label{chi0iaia2omega}
\end{eqnarray}
and, for the virtual-occupied/virtual-occupied block, $\chi_{0,ai,ai}(\omega',\omega) = \chi_{0,ia,ia}(\omega',-\omega)$ and  $\chi_{0,ai,ai}(\omega) = \chi_{0,ia,ia}(-\omega)$.
The matrix elements of the BSE2 correlation kernel are then found straightforwardly by doing the matrix multiplications and contour-integrating over the frequencies in the upper-half complex plane in Eq.~(\ref{eq:fcBSE2}). For the matrix elements in the occupied-virtual/occupied-virtual (ov/ov) block (contributing to the linear-response matrix usually denoted by $\mA$), we find
\begin{equation}
  \begin{split}
    f_{\tc,\text{BSE2},ia,jb}(\omega) =&  
- \sum_{kc} \dfrac{\langle jc||ik \rangle \langle ka||cb \rangle}{\omega -(\varepsilon_b+\varepsilon_c-\varepsilon_i-\varepsilon_k) }
\\
&
- \sum_{kc} \dfrac{\langle jk||ic \rangle \langle ca||kb \rangle} {\omega-(\varepsilon_a + \varepsilon_c - \varepsilon_j - \varepsilon_k)}
\\
&+ \dfrac{1}{2} \sum_{kl}\dfrac{ \langle aj||kl\rangle\langle lk||bi \rangle}{\omega - (\varepsilon_a+\varepsilon_b - \varepsilon_k-\varepsilon_l)  } 
\\
&+ \dfrac{1}{2} \sum_{cd} \dfrac{\langle aj||cd\rangle\langle dc||bi \rangle }{\omega - (\varepsilon_c+\varepsilon_d - \varepsilon_i-\varepsilon_j) }. 
\label{eq:eff kernel iajb}
  \end{split}
\end{equation}
Note that the denominators of Eq.~(\ref{eq:eff kernel iajb}) contain the sum of two virtual spin-orbital energies minus the sum of two occupied spin-orbital energies, i.e. a non-interacting double-excitation energy. Thus, the denominators are small (and therefore the kernel can be large) whenever $\omega$ is close to a non-interacting double-excitation energy. The matrix elements in Eq.~(\ref{eq:eff kernel iajb}) are identical (at least for real-valued spin orbitals) to the kernel matrix elements recently derived by Zhang \emph{et al.}~\cite{ZhaSteYan-JCP-13} in the TDA~\cite{RebTou-JJJ-XX-note1}. The matrix elements in Eq.~(\ref{eq:eff kernel iajb}) also show some similitude with the SOPPA kernel~\cite{OddJor-JCP-77,NieJorOdd-JCP-80,HuiCas-ARX-10,Hui-THESIS-11} and the second RPA kernel~\cite{SanRomOniMar-JCP-11}. Similarly, for the matrix elements of the BSE2 correlation kernel in the occupied-virtual/virtual-occupied (ov/vo) block (contributing to the linear-response matrix usually denoted by $\mB$), we find
\begin{equation}
  \begin{split}
    f_{\tc,\text{BSE2},ia,bj} = & 
- \sum_{kc} \dfrac{\langle bc||ik \rangle \langle ka||cj \rangle}{ - (\varepsilon_b + \varepsilon_c - \varepsilon_i-\varepsilon_k)}
\\ & 
-\sum_{kc} \dfrac{\langle bk||ic \rangle \langle ca||kj \rangle} {- (\varepsilon_a + \varepsilon_c - \varepsilon_j - \varepsilon_k) }
\\
    & + \dfrac{1}{2} \sum_{kl}\dfrac{ \langle ab||kl\rangle \langle
      lk||ji \rangle}{-(\varepsilon_a+\varepsilon_b
      -\varepsilon_k-\varepsilon_l) } \\ & + \dfrac{1}{2}
    \sum_{cd} \dfrac{ \langle ab||cd\rangle \langle dc||ji \rangle}{
      -(\varepsilon_c+\varepsilon_d- \varepsilon_i-\varepsilon_j) 
      } ,
      \label{eq:eff kernel iabj}
  \end{split}
\end{equation}
which turn out to be independent of the frequency. To the best of our knowledge, the matrix elements in Eq.~(\ref{eq:eff kernel iabj}) had never been given in the literature before. It is easy to check that the ov/ov block is Hermitian, $f_{\tc,\text{BSE2},ia,jb}(\omega) = f_{\tc,\text{BSE2},jb,ia}(\omega)^*$, and that the ov/vo block is symmetric, $f_{\tc,\text{BSE2},ia,bj} = f_{\tc,\text{BSE2},jb,ai}$.

The matrix elements of the BSE2 correlation kernel display sums over either one occupied and one virtual orbital (for the $\ph$ terms) or over two
occupied or two virtual orbitals (for the $\pp$ terms). In a straightforward implementation, the computational cost of the latter scales as $N_\text{o}^2N_\text{v}^4$ where $N_\text{o}$ is the number of occupied orbitals and $N_\text{v}$ the number of virtual ones. However, in the case of the long-range interaction, the computational cost of the BSE2 correlation kernel could be made low, e.g. by approximating the long-range two-electron integrals by multipole expansions~\cite{HetSchStoWer-JCP-00}.

\section{Practical resolution and computational details}
\label{sec:RSH_BSE}

\subsection{Perturbative resolution}

In the range-separated scheme that we propose, we approximate the inverse of the linear-response function as [combining Eqs.~(\ref{eq:chim1-RSTDDFT}) and~(\ref{eq:chim1-RSH-BSE2})]
\begin{equation}
\chi^{-1}(\omega) \approx \chi_0^{-1}(\omega) - f_{\Hx,\HF}^{\lr} - f_{\Hxc}^{\sr} - f_{\c,\text{BSE2}}^{\lr} (\omega),
\label{chim1RS}
\end{equation}
where $\chi_0(\omega)$ is the RSH non-interacting linear-response function and $f_{\c,\text{BSE2}}^{\lr}(\omega)$ is the BSE2 correlation kernel for the long-range electron-electron interaction. We note that, according to Eq.~(\ref{eq:effectiveBSEinv}), instead of $\chi_0^{-1}(\omega)$, we should use in Eq.~(\ref{chim1RS}) the inverse of the long-range IP linear-response function $(\chi^{\lr}_\IP)^{-1}(\omega)$ constructed with the long-range interacting Green function. This could be accounted for by either adding quasiparticle corrections to the orbital energies, as done in Ref.~\onlinecite{ZhaSteYan-JCP-13}, or adding self-energy contributions to the long-range correlation kernel~\cite{SanRomOniMar-JCP-11}. These contributions can generally be important when using HF orbitals or DFT orbitals with semilocal DFAs. However, in the case of range separation, the orbital energies obtained with long-range HF exchange are already good approximations to quasiparticle energies~\cite{TsuSonSuzHir-JCP-10,KroSteRefBae-JCTC-12}. It is thus reasonable to use the approximation $(\chi^{\lr}_\IP)^{-1}(\omega) \approx \chi_0^{-1}(\omega)$. We come back to the possibility of adding quasiparticle corrections in Section~\label{sec:lrexcit} and discuss their effects on He, Be, and H$_2$ in Section~\ref{sec:results_small}.

When projected in the basis of the RSH spin orbitals, Eq.~(\ref{chim1RS}) leads to the self-consistent pseudo-Hermitian eigenvalue equation
\begin{equation}
  \left(
  \begin{array}{cc}
    \mA(\omega_n) & \mB \\
    \mB^* & \mA(-\omega_n)^*
  \end{array}
  \right) 
  \left(
  \begin{array}{c}
    \mX_n \\
    \mY_n\\
  \end{array}
  \right) 
  = 
  {\omega_n}
  \left(
  \begin{array}{cc}
    \bm{1} & \bm{0} \\
    \bm{0} & \bm{-1}
  \end{array}
  \right) 
  \left(
  \begin{array}{c}
    \mX_n \\
    \mY_n\\
  \end{array}
  \right),
\label{eq:eigenvalueeq}
\end{equation}
where $\omega_n$ are the excitation (or diexcitation) energies, $(\mX_n,\mY_n)$ are the associated linear-response eigenvectors, and the matrix elements of $\mA$ and $\mB$ are given by
\begin{eqnarray}
    A_{ia,jb}(\omega) &=& (\varepsilon_{a} - \varepsilon_{i}) \delta_{ij} \delta_{ab} +  \langle a j | \mv | i b \rangle - \langle a j  | \mv^{\lr} | b i \rangle
\nonumber\\
&& \qquad + f_{\x\tc,ia,jb}^{\sr} + f^{\lr}_{\tc,\text{BSE2},ia,jb}(\omega),
\label{ARSHBSE2}
\end{eqnarray}
and
\begin{eqnarray}
B_{ia,jb} &=& \langle a b | \mv |i j \rangle - \langle a b |\mv^{\lr} |j i \rangle 
\nonumber\\
&& \qquad  + f_{\x\tc,ia,bj}^{\sr} + f^{\lr}_{\tc,\text{BSE2},ia,bj},
\label{BRSHBSE2}
\end{eqnarray}
where $\varepsilon_{p}$ are the RSH spin-orbital energies, $\langle p q | \mv|r s \rangle$ and $\langle p q | \mv^{\lr} |r s \rangle$ are two-electron integrals in the RSH spin-orbital basis associated with the Coulomb interaction $\mv$ and the long-range interaction $\mv^{\lr}$, respectively, and $f_{\x\tc,pq,rs}^{\sr}$ are the matrix elements of the short-range exchange-correlation kernel. The matrix elements of the long-range BSE2 correlation kernel $f^{\lr}_{\tc,\text{BSE2},pq,rs}$ are given in Eqs.~(\ref{eq:eff kernel iajb}) and~(\ref{eq:eff kernel iabj}) using in these expressions long-range two-electron integrals $\langle p q || r s \rangle \to \langle p q |\mv^{\lr}| r s \rangle - \langle p q |\mv^{\lr}| s r \rangle$ and RSH spin-orbital energies $\varepsilon_{p}$.

The resolution of the self-consistent eigenvalue equation~(\ref{eq:eigenvalueeq}) is more complicated than in the standard case of a frequency-independent matrix $\b{A}$. Following Zhang \emph{et al.}~\cite{ZhaSteYan-JCP-13}, for a first exploration of the method, we work within the TDA (i.e., we set $\b{B}=\b{0}$) and use a non-self-consistent perturbative resolution. We thus decompose the matrix $\b{A}$ in Eq.~(\ref{ARSHBSE2}) as the sum of the frequency-independent RSH contribution~\cite{RebSavTou-MP-13} and the long-range frequency-dependent BSE2 correlation kernel contribution
\begin{equation}
\mA(\omega) = \mA_\text{RSH} + \b{f}_{\tc,\text{BSE2}}^{\lr}(\omega).
\end{equation}
The TDRSH linear-response equation is first solved in the TDA,
\begin{equation}
  \mA_{\text{RSH}} \mX_{0,n} = \omega_{0,n} \mX_{0,n},
  \label{eq:omega_n TDA}
\end{equation}
where $\omega_{0,n}$ and $\mX_{0,n}$ are the corresponding excitation energies and linear-response eigenvectors, respectively. The effect of the long-range BSE2 correlation kernel is then added perturbatively to obtain the excitation energies
\begin{equation}
  \omega_n = \omega_{0,n} + Z_n \, \mX_{0,n}^{\dagger}
  \, \b{f}_{\tc,\text{BSE2}}^{\lr}(\omega_{0,n}) \, \mX_{0,n},
  \label{eq:omega_n pert}
\end{equation}
where $Z_n$ is the normalization factor
\begin{equation}
  Z_n = \left(1 - \mX_{0,n}^{\dagger} \left.\dfrac{\partial
    \b{f}_{\tc,\text{BSE2}}^{\lr}(\omega)}{\partial \omega}\right|_{\omega=
    \omega_{0,n}} \mX_{0,n}\right)^{-1}.
\end{equation}
As pointed out by Zhang \emph{et al.}~\cite{ZhaSteYan-JCP-13}, the effect of the normalization factor $Z_n$ turns out to be very small ($Z_n$ is always very close to 1, especially in the range-separated case), but we keep it in our calculations. We note that the expression of the correction $\mX_{0,n}^{\dagger} \, \b{f}_{\tc,\text{BSE2}}^{\lr}(\omega_{0,n}) \, \mX_{0,n}$ in Eq.~(\ref{eq:omega_n pert}) is very similar (but not identical) to the so-called ``direct'' contribution of the CIS(D) correction~\cite{HeaRicOumlee-CPL-94,RheHea-JPCA-07}. As for CIS(D), it is easy to check that $\mX_{0,n}^{\dagger} \, \b{f}_{\tc,\text{BSE2}}^{\lr}(\omega_{0,n}) \, \mX_{0,n}$ contains only connected terms and thus provides a size-consistent correction to the excitation energies.
Using this non-self-consistent perturbative resolution has the consequence that the total number of calculated excitation energies is equal to the number of single excitations, so we cannot obtain excitations with primarily double-excitation character. However, the BSE2 correlation kernel brings the effects of non-interacting double excitations on excited states with dominant single-excitation character.

The method defined by Eqs.~(\ref{eq:omega_n TDA}) and~(\ref{eq:omega_n pert}) will be referred to as TDRSH+BSE2. When the range-separation parameter $\mu$ is set to zero, all long-range contributions vanish, and it reduces to the standard time-dependent Kohn-Sham (TDKS) method in the TDA. When $\mu$ goes to $+\infty$, all short-range contributions vanish, and it reduces to time-dependent Hartree-Fock (TDHF) within the TDA [i.e., configuration-interaction singles (CIS)] with a BSE2 correction, which will be referred to as TDHF+BSE2. As regards the density-functional approximation, in this work, we use the short-range LDA exchange-correlation functional of Ref.~\onlinecite{PazMorGorBac-PRB-06} in the ground-state RSH calculations (i.e., for determining the RSH orbitals and orbital energies) and the corresponding short-range LDA exchange-correlation kernel~\cite{RebSavTou-MP-13} in the linear-response TDRSH calculations.

\subsection{Long-range excitation energies}
\label{sec:lrexcit}

For He, Be, and H$_2$, we also perform calculations of long-range excitation energies as a function of $\mu$ (i.e., along the range-separated adiabatic connection, similarly to Refs.~\onlinecite{RebTouTeaHelSav-JCP-14,RebTouTeaHelSav-MP-15,RebTouTeaHelSav-PRA-15}) obtained by removing the contribution from the short-range Hxc kernel $f_{\Hxc}^\sr$ in the matrix elements $A_{ia,jb}$ of Eq.~(\ref{ARSHBSE2}), i.e.
\begin{eqnarray}
    A_{ia,jb}^\lr(\omega) &=& (\varepsilon_{a} - \varepsilon_{i}) \delta_{ij} \delta_{ab} +  \langle a j | \mv^\lr | i b \rangle - \langle a j  | \mv^{\lr} | b i \rangle
\nonumber\\
&& \qquad + f^{\lr}_{\tc,\text{BSE2},ia,jb}(\omega),
\label{AlrRSHBSE2}
\end{eqnarray}
within the perturbative resolution of Eqs.~(\ref{eq:omega_n TDA}) and~(\ref{eq:omega_n pert}) in the TDA. The orbitals and orbital energies used in Eq.~(\ref{AlrRSHBSE2}) are still the RSH ones (i.e., with the short-range LDA exchange-correlation functional), as for the other calculations. The obtained long-range excitation energies are approximations to the excitation energies of the long-range interacting Hamiltonian of Eq.~(\ref{Hlr}), which reduces to the LDA orbital energy differences at $\mu=0$ and to the TDHF+BSE2 excitation energies for $\mu\to\infty$. These long-range excitation energies allows us to test the effect of the BSE2 correlation kernel independently of the approximation used for the short-range exchange-correlation kernel, since we have accurate reference values for these quantities from Ref.~\onlinecite{RebTouTeaHelSav-JCP-14}.

For these systems, we also test the addition of the perturbative quasiparticle correction using the long-range second-order correlation self-energy, similarly to Ref.~\onlinecite{ZhaSteYan-JCP-13}, i.e. replacing the RSH orbital energies $\varepsilon_{p}$ in Eq.~(\ref{AlrRSHBSE2}), including in the long-range BSE2 correlation kernel $f^{\lr}_{\tc,\text{BSE2},ia,jb}(\omega)$, by the quasiparticle energies
\begin{eqnarray}
\tilde{\varepsilon}_{p} = \varepsilon_{p} + z_p \; \Sigma_{\c,p p}^\lr(\varepsilon_{p}),
\label{tildeeps}
\end{eqnarray}
with the renormalization factor $z_p = [1 - (\partial \Sigma_{\c,p p}^\lr(\omega)/ \partial \omega)_{\omega=\varepsilon_{p}} ]^{-1}$. In Eq.~(\ref{tildeeps}), $\Sigma_{\c,p p}^\lr(\varepsilon_{p})$ is the diagonal matrix element of the frequency-dependent long-range second-order correlation self-energy $\Sigma_{\c}^\lr(\omega)$ over the RSH spin orbital $\varphi_p(\b{x})$ evaluated at $\omega=\varepsilon_{p}$, whose expression is
\begin{eqnarray}
\Sigma_{\c,p p}^\lr(\omega) &=& \frac{1}{2} \sum_{iab} \frac{|\langle a b |\mv^{\lr} |p i \rangle - \langle a b |\mv^{\lr} |i p \rangle|^2}{\omega + \varepsilon_i -  \varepsilon_a - \varepsilon_b}
\nonumber\\
&& + \frac{1}{2} \sum_{ija} \frac{|\langle i j |\mv^{\lr} |p a \rangle - \langle i j |\mv^{\lr} |a p \rangle|^2}{\omega + \varepsilon_a -  \varepsilon_i - \varepsilon_j},
\label{Sigmacpplr}
\end{eqnarray}
where $i,j$ and $a,b$ refer to occupied and virtual RSH spin orbitals, respectively. This quasiparticle correction will be denoted by GW2 since it is a second-order $GW$-type correction. The resulting method will thus be referred to as GW2+TDHF+BSE2.

\subsection{Computational details}
\label{sec:computational dyn}

We calculate vertical excitation energies of four small molecules, N$_2$, CO, H$_2$CO, and C$_2$H$_4$, at their experimental geometries~\cite{Huber1979,LeF-MP-91,GurVeyAlc-book-89,Her-BOOK-66}, using the Sadlej+ basis sets~\cite{Casida1998}. Our reference values are obtained by equation-of-motion coupled-cluster singles doubles (EOM-CCSD) calculations performed with GAUSSIAN 09~\cite{Gaussian-PROG-09}. For each molecule, we report the first 14 excited states found with the EOM-CCSD method. For each molecule, we perform a self-consistent ground-state RSH calculation using the short-range LDA exchange-correlation functional of Ref.~\onlinecite{PazMorGorBac-PRB-06}, followed by a spin-adapted closed-shell TDRSH linear-response calculation in the TDA using the short-range LDA exchange-correlation kernel~\cite{RebSavTou-MP-13}, as implemented in a development version of MOLPRO~\cite{Molproshort-PROG-12}. The TDRSH+BSE2 excitation energies are then calculated by a spin-adapted closed-shell version of Eq.~(\ref{eq:omega_n pert}) implemented in a homemade software interfaced with MOLPRO (see Ref.~\onlinecite{Reb-THESIS-14_eng} for details). The range-separation parameter $\mu$ is set to 0.35 bohr$^{-1}$ which yields a minimal mean absolute deviation (MAD) over the four molecules of the TDRSH+BSE2 excitation energies with respect to the EOM-CCSD references. We note that it has been proposed to adjust the value of $\mu$ for each system by imposing a self-consistent Koopmans' theorem condition~\cite{SteKroBae-JACS-09,SteKroBae-JCP-09} or, equivalently, minimizing the deviation from the piecewise linearity behavior of the total energy as a function of the electron number~\cite{SteAutGovKroBae-JPCL-12,GlePeaToz-JCTC-13}. This approach is appealing but it has the disadvantage of being non size consistent~\cite{KarKroKum-JCP-13}, so we prefer to use a fixed value of $\mu$, independent of the system. For comparison, we also perform standard, linear-response TDKS calculations with the LDA functional~\cite{PerWan-PRB-92}, as well as TDHF and TDHF+BSE2 calculations, all in the TDA. In the TDA, $X_{0,n,ia}$ can be considered as the coefficient of the (spin-orbital) single excitation $i\to a$ in the wave function of the excited state $n$. Each excited state was thus assigned by looking at its symmetry and at the leading orbital contributions to the excitation. 

The calculations of the long-range excitation energies for He, Be, and H$_2$ are done similarly except that the short-range LDA exchange-correlation kernel is removed in the TDRSH linear-response calculation. The GW2 quasiparticle correction is calculated using a spin-adapted closed-shell version of Eq.~(\ref{tildeeps}). We use an uncontracted t-aug-cc-pV5Z basis set for He, an uncontracted d-aug-cc-pVDZ basis set for Be, and an uncontracted d-aug-cc-pVTZ basis set for H$_2$, for which we have reference long-range excitation energies obtained at the full configuration-interaction (FCI) level using an accurate Lieb-optimized short-range potential~\cite{RebTouTeaHelSav-JCP-14}.

\section{Results and discussion}
\label{sec:results dyn}

\subsection{Long-range excitation energies of the He and Be atoms and of the H$_2$ molecule}
\label{sec:results_small}

\begin{figure}
\includegraphics[scale=0.30,angle=-90]{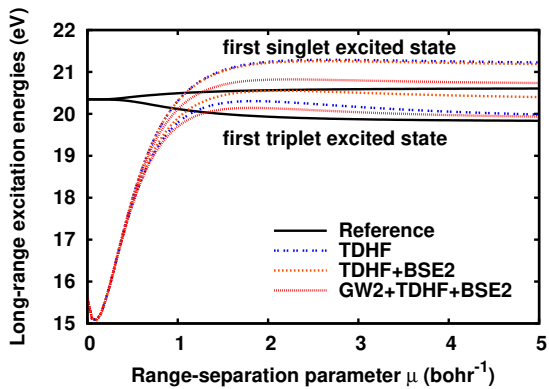}
\caption{Long-range excitation energies to the first triplet and singlet excited states of the He atom as a function of the range-separation parameter $\mu$, obtained by long-range TDHF, long-range TDHF+BSE2, and long-range GW2+TDHF+BSE2 calculations in the TDA using RSH (with the short-range LDA functional) orbitals and an uncontracted t-aug-cc-pV5Z basis set. The reference FCI long-range excitation energies are from Ref.~\onlinecite{RebTouTeaHelSav-JCP-14}.
}
\label{fig:he}
\end{figure}

The long-range excitation energies to the first triplet and singlet excited states of the He atom are plotted as a function of the range-separation parameter $\mu$ in Figure~\ref{fig:he}. The triplet and singlet excitation energies are identical at $\mu=0$, where they reduce to the non-interacting Kohn-Sham excitation energies. When increasing $\mu$, i.e. when adding the long-range interaction, this degeneracy is lifted and the excitation energies tend to the physical excitation energies in the limit $\mu\to\infty$. At $\mu=0$, for all the approximate methods tested here, the long-range excitation energies reduce to LDA orbital energy differences, which, as well known for Rydberg states, strongly underestimate the exact Kohn-Sham orbital energy differences (by about 5 eV in the present case). This underestimation of the long-range excitation energies is progressively eliminated by increasing the value of $\mu$ until $\mu \approx 1$ bohr$^{-1}$. For $\mu \gtrsim 1.5$ bohr$^{-1}$, with all the approximate methods, the long-range excitation energies vary much less and are a bit too high compared to the reference FCI long-range excitation energies. The BSE2 correlation kernel has almost no effect for the singlet excited state, while it increases the excitation energy for the triplet excited state which leads to a larger error at large $\mu$. The GW2 quasiparticle correction systematically decreases the excitation energies, leading to smaller errors at large $\mu$ for both singlet and triplet excitation energies. The GW2 correction on the excitation energies is relatively large (0.5 eV) for large $\mu$, but decreases when $\mu$ is decreased, being less than 0.2 eV for $\mu \leq 1$ bohr$^{-1}$ and about 0.01 eV for $\mu=0.35$ bohr$^{-1}$ (the value of $\mu$ used for the other systems in Section~\ref{sec:results_more}).
 
\begin{figure}
\includegraphics[scale=0.30,angle=-90]{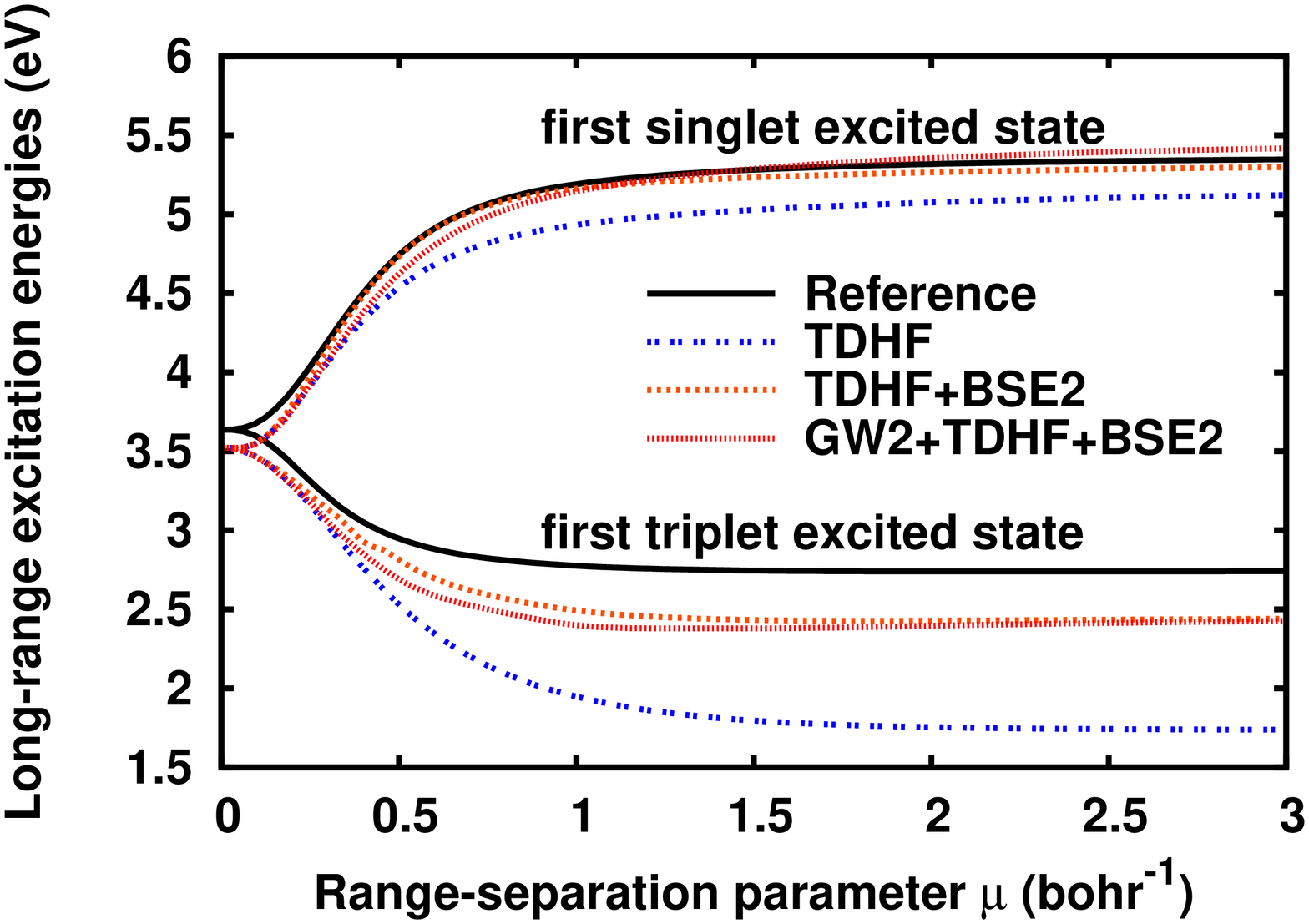}
\caption{Long-range excitation energies to the first triplet and singlet excited states of the Be atom as a function of the range-separation parameter $\mu$, obtained by long-range TDHF, long-range TDHF+BSE2, and long-range GW2+TDHF+BSE2 calculations in the TDA using RSH (with the short-range LDA functional) orbitals and an uncontracted d-aug-cc-pVDZ basis set. The reference FCI long-range excitation energies are from Ref.~\onlinecite{RebTouTeaHelSav-JCP-14}.
}
\label{fig:be}
\end{figure}

\begin{figure}
\includegraphics[scale=0.30,angle=-90]{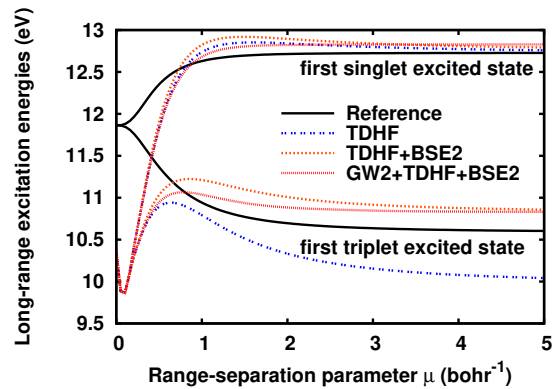}
\caption{Long-range excitation energies to the first triplet and singlet excited states of the H$_2$ molecule at the equilibrium internuclear distance as a function of the range-separation parameter $\mu$, obtained by long-range TDHF, long-range TDHF+BSE2, and long-range GW2+TDHF+BSE2 calculations in the TDA using RSH (with the short-range LDA functional) orbitals and an uncontracted d-aug-cc-pVTZ basis set. The reference FCI long-range excitation energies are from Ref.~\onlinecite{RebTouTeaHelSav-JCP-14}.
}
\label{fig:h2_eq}
\end{figure}

\begingroup
\squeezetable
\begin{table*}
  \centering
  \caption
    {Excitation energies of N$_2$ calculated by linear-response TDKS (with the LDA functional), TDRSH and TDRSH+BSE2 (with the short-range LDA functional and $\mu=0.35$ bohr$^{-1}$), TDHF and TDHF+BSE2, all within the TDA. The EOM-CCSD excitation energies are taken as reference. The Sadlej+ basis set is used.}
  \begin{tabular}{c l c c H c c H c c}
    \hline\hline
State 	 			& Transition		& TDKS	& TDRSH	& TDRSH+BSE2' & TDRSH+BSE2 & TDHF	& TDHF+BSE2' & TDHF+BSE2 & EOM-CCSD \\
\hline
\multicolumn{10}{c}{ Valence excitation energies (eV)} \\
\hline
$^3\Sigma_\text{u}^+$  & $1\pi_\text{u} \rightarrow 1\pi_\text{g}$           & 8.08	& 7.74   & 7.93  & 7.93 & 6.23	& 8.76	& 8.88	& 7.72\\
$^3 \Pi_\text{g}$      & $3\sigma_\text{g} \rightarrow 1\pi_\text{g}$        & 7.58	& 7.85   & 8.05  & 8.05 & 7.99	& 10.80	& 10.97	& 8.16\\
$^3\Delta_\text{u}$    & $1\pi_\text{u} \rightarrow 1\pi_\text{g}$           & 8.88	& 8.54   & 8.73  & 8.74 & 7.32	& 9.84	& 9.96	& 9.07\\
$^1\Pi_\text{g}$       & $3\sigma_\text{g} \rightarrow 1\pi_\text{g}$        & 9.17	& 9.50   & 9.68  & 9.68 & 10.02 & 12.31	& 12.43	& 9.55\\
$^3\Sigma_\text{u}^-$  & $1\pi_\text{u} \rightarrow 1\pi_\text{g}$           & 9.65	& 9.34   & 9.53  & 9.53 & 8.50	& 10.67	& 10.77	& 10.00\\
$^1\Sigma_\text{u}^-$  & $1\pi_\text{u} \rightarrow 1\pi_\text{g}$           & 9.65	& 9.34   & 9.53  & 9.53 & 8.50	& 10.73	& 10.84	& 10.24\\
$^1\Delta_\text{u}$    & $1\pi_\text{u} \rightarrow 1\pi_\text{g}$           & 10.25	& 9.98   & 10.18 & 10.18& 9.06	& 11.20	& 11.30	& 10.66\\
$^3\Pi_\text{u}$       & $2\sigma_\text{u} \rightarrow 1\pi_\text{g}$        & 10.42	& 10.77  & 10.97 & 10.97& 11.74 & 14.63	& 14.82	& 11.36\\
\hline 
\multicolumn{10}{c}{ Rydberg excitation energies (eV)} \\                                 
\hline 
$^3\Sigma_\text{g}^+$  &$3\sigma_\text{g} \rightarrow 4\sigma_\text{g}$      & 10.28  & 11.47   & 11.56 & 11.56 & 13.12	& 13.93	& 13.94	& 11.74\\
$^1\Sigma_\text{g}^+$  & $3\sigma_\text{g} \rightarrow 4\sigma_\text{g}$     & 10.40  & 11.94   & 11.98 & 11.98 & 14.01	& 14.22	& 14.22	& 12.15\\
$^3\Sigma_\text{u}^+$  & $3\sigma_\text{g} \rightarrow 3\sigma_\text{u}$     & 10.63	& 12.30   & 12.40 & 12.40 & 14.21 & 15.05	& 15.07	& 12.70\\
$^3\Pi_\text{u}$       & $3\sigma_\text{g} \rightarrow 2\pi_\text{u}$        & 10.99  & 12.30   & 12.36 & 12.36 & 13.04	& 13.42	& 13.43	& 12.71\\
$^1\Pi_\text{u}$       & $3\sigma_\text{g} \rightarrow 2\pi_\text{u}$        & 10.98  & 12.39   & 12.44 & 12.44 & 13.23	& 13.45	& 13.45	& 12.77\\
$^1\Sigma_\text{u}^+$  & $3\sigma_\text{g} \rightarrow 3\sigma_\text{u}$     & 10.62	& 12.43   & 12.51 & 12.51  & 14.31	& 15.02	& 15.04	& 12.82\\
\hline
\multicolumn{10}{c}{ Ionization threshold: $-\epsilon_{\text{HOMO}}$ (eV)} \\
\hline
 & & 6.30 & 14.94 &       &       & 16.74 &       &       & \\
\hline
\multicolumn{10}{c}{MAD of excitation energies with respect to EOM-CCSD (eV)} \\
\hline
Valence & & 0.48 & 0.47 & 0.35 & 0.35 & 1.14 & 1.52 & 1.65 & - \\
Rydberg & & 1.83 & 0.34 & 0.27 & 0.27 & 1.17 & 1.70 & 1.71 & - \\
Total   & & 1.06 & 0.41 & 0.32 & 0.32 & 1.15 & 1.60 & 1.68 & - \\
\hline     
\multicolumn{10}{c}{Maximum absolute deviation of excitation energies with respect to EOM-CCSD (eV)} \\
\hline
 & & 2.19 & 0.90 & 0.71 & 0.71 & 1.86 & 3.28 & 3.47 & - \\
\hline\hline
  \end{tabular}
    \label{N2table_dyn}
\end{table*}
\endgroup
\begingroup
\squeezetable
\begin{table*}
  \centering
   \caption
    {Same as Table~\ref{N2table_dyn} for CO.}
  \begin{tabular}{c l c c H c c H c c}
    \hline\hline
State 	 			& Transition				& TDKS	& TDRSH	& TDRSH+BSE2' & TDRSH+BSE2 & TDHF	& TDHF+BSE2' & TDHF+BSE2 & EOM-CCSD \\
\hline
\multicolumn{10}{c}{ Valence excitation energies (eV)} \\
\hline
$^3\Pi$                 & $5a_1(\sigma) \rightarrow 2e_1(\pi^*)$         & 6.04    & 6.10  & 6.32  & 6.32  & 5.85  & 8.15  & 8.27  & 6.45 \\
$^3\Sigma^+$            & $1e_1(\pi) \rightarrow 2e_1(\pi^*)$            & 8.54    & 8.45  & 8.63  & 8.63  & 7.79  & 10.26 & 10.38 & 8.42 \\
$^1\Pi$                 & $5a_1(\sigma) \rightarrow 2e_1(\pi^*)$         & 8.42    & 8.68  & 8.88  & 8.88  & 9.08  & 10.86 & 10.94 & 8.76 \\
$^3\Delta$              & $1e_1(\pi) \rightarrow 2e_1(\pi^*)$            & 9.20    & 9.13  & 9.31  & 9.31  & 8.74  & 11.08 & 11.19 & 9.39 \\
$^3\Sigma^-$            & $1e_1(\pi) \rightarrow 2e_1(\pi^*)$            & 9.84    & 9.80  & 9.97  & 9.98  & 9.73  & 11.68 & 11.76 & 9.97 \\
$^1\Sigma^-$            & $1e_1(\pi) \rightarrow 2e_1(\pi^*)$            & 9.84    & 9.80  & 9.98  & 9.98  & 9.73  & 11.73 & 11.82 & 10.19\\
$^1\Delta$              & $1e_1(\pi) \rightarrow 2e_1(\pi^*)$            & 10.33   & 10.32 & 10.50 & 10.50 & 10.15 & 11.98 & 12.05 & 10.31\\
$^3\Pi$                 & $4a_1(\sigma) \rightarrow 2e_1(\pi^*)$         & 11.43   & 11.96 & 12.12 & 12.12 & 13.31 & 15.59 & 15.70 & 12.49\\
\hline
\multicolumn{10}{c}{Rydberg excitation energies (eV)} \\
\hline
$^3\Sigma^+$            & $5a_1(\sigma) \rightarrow 6a_1(\sigma)$       & 9.56    & 10.34 & 10.46 & 10.46 & 11.18 & 12.07 & 12.09 & 10.60\\
$^1\Sigma^+$            & $5a_1(\sigma) \rightarrow 6a_1(\sigma)$       & 9.95    & 11.12 & 11.20 & 11.20 & 12.27 & 12.61 & 12.61 & 11.15\\
$^3\Sigma^+$            & $5a_1(\sigma) \rightarrow 7a_1(\sigma)$       & 10.26   & 11.08 & 11.17 & 11.17 & 12.42 & 12.82 & 12.83 & 11.42\\
$^1\Sigma^+$            & $5a_1(\sigma) \rightarrow 7a_1(\sigma)$       & 10.50   & 11.30 & 11.38 & 11.38 & 12.79 & 12.91 & 12.91 & 11.64\\
$^3\Pi$                 & $5a_1(\sigma) \rightarrow 3e_1(\pi)$          & 10.39   & 11.26 & 11.34 & 11.34 & 12.60 & 13.19 & 13.20 & 11.66\\
$^1\Pi$                 & $5a_1(\sigma) \rightarrow 3e_1(\pi)$          & 10.50   & 11.45 & 11.52 & 11.52 & 12.88 & 13.21 & 13.21 & 11.84\\
\hline
\multicolumn{10}{c}{ Ionization threshold: $-\epsilon_{\text{HOMO}}$ (eV)} \\
\hline
 & & 9.12 & 13.49 &       &       & 15.11 &       &       & \\
\hline
\multicolumn{10}{c}{ MAD of excitation energies with respect to the EOM-CCSD calculation (eV)} \\
\hline
Valence & & 0.33 & 0.23 & 0.16 & 0.16 & 0.49 & 1.92 & 2.02 & - \\
Rydberg & & 1.19 & 0.29 & 0.22 & 0.22 & 0.97 & 1.42 & 1.42 & - \\
Total   & & 0.70 & 0.26 & 0.19 & 0.19 & 0.69 & 1.70 & 1.76 & - \\
\hline
\multicolumn{10}{c}{Maximum absolute deviation of excitation energies with respect to EOM-CCSD (eV)} \\
\hline
 & & 1.34 & 0.53 & 0.37 & 0.36 & 1.16 & 3.10 & 3.22 & - \\
\hline\hline
  \end{tabular}
    \label{COtable_dyn}
\end{table*}
\endgroup

\begingroup
\squeezetable
\begin{table*}
  \centering
  \caption
    {Same as Table~\ref{N2table_dyn} for H$_2$CO.}
  \begin{tabular}{c l c c H c c H c c}
    \hline\hline
State 	 			& Transition                    & TDKS    & TDRSH	& TDRSH+BSE2' & TDRSH+BSE2 & TDHF	& TDHF+BSE2' & TDHF+BSE2 & EOM-CCSD \\
\hline \hline
\multicolumn{10}{c}{ Valence excitation energies (eV)} \\
\hline
$^3A_2$         & $2b_2(n) \rightarrow 2b_1(\pi^*)$             & 3.08  & 3.17  & 3.45  & 3.45  & 3.76  & 6.66  & 6.86  & 3.56 \\
$^1A_2$         & $2b_2(n)\rightarrow 2b_1(\pi^*)$              & 3.70  & 3.82  & 4.11  & 4.11  & 4.58  & 7.20  & 7.37  & 4.03 \\
$^3A_1$         & $1b_1(\pi)\rightarrow 2b_1(\pi^*)$            & 6.35  & 6.08  & 6.39  & 6.39  & 4.96  & 8.08  & 8.30  & 6.06 \\
$^3B_1$         & $5a_1(\sigma)\rightarrow 2b_1(\pi^*)$         & 7.77  & 8.09  & 8.39  & 8.40  & 8.60  & 12.01 & 12.28 & 8.54 \\
\hline
\multicolumn{10}{c}{Rydberg excitation energies (eV)} \\
\hline
$^3B_2$         & $2b_2(n)\rightarrow 6a_1(\sigma)$             & 5.85  & 6.83  & 6.92  & 6.92  & 8.17  & 8.63  & 8.63  & 6.83\\
$^1B_2$         & $2b_2(n)\rightarrow 6a_1(\sigma)$             & 5.93  & 7.01  & 7.07  & 7.08  & 8.56  & 8.72  & 8.72  & 7.00\\
$^3B_2$         & $2b_2(n)\rightarrow 7a_1(\sigma)$             & 6.96  & 7.69  & 7.81  & 7.81  & 9.04  & 9.83  & 9.85  & 7.73\\
$^3A_1$         & $2b_2(n) \rightarrow 3b_2(\sigma)$            & 6.73  & 7.77  & 7.83  & 7.83  & 9.24  & 9.58  & 9.58  & 7.87\\
$^1B_2$         & $2b_2(n)\rightarrow 7a_1(\sigma)$             & 7.04  & 7.91  & 8.00  & 8.00  & 9.41  & 9.78  & 9.78  & 7.93\\
$^1A_1$         & $2b_2(n) \rightarrow 3b_2(\sigma)$            & 6.78  & 7.93  & 7.97  & 7.97  & 9.53  & 10.00 & 10.01 & 7.99\\
$^1A_2$         & $2b_2(n) \rightarrow 3b_1(\pi)$               & 7.55  & 8.32  & 8.39  & 8.39  & 10.04 & 10.26 & 10.26 & 8.45\\
$^3A_2$         & $2b_2(n) \rightarrow 3b_1(\pi)$               & 7.58  & 8.31  & 8.38  & 8.38  & 9.93  & 11.04 & 11.07 & 8.47\\
$^3B_2$         & $2b_2(n)\rightarrow 8a_1(\sigma)$             & 7.97  & 8.90  & 8.98  & 8.98  & 10.21 & 11.89 & 11.96 & 8.97\\
$^1B_2$         & $2b_2(n)\rightarrow 8a_1(\sigma)$             & 8.19  & 9.17  & 9.25  & 9.25  & 10.86 & 11.05 & 11.05 & 9.27\\
\hline
\multicolumn{10}{c}{ Ionization threshold: $-\epsilon_{\text{HOMO}}$ (eV)} \\
\hline
 & & 6.30 & 10.33 &       &       & 12.04 &       &       & \\
\hline
\multicolumn{10}{c}{ MAD of excitation energies with respect to the EOM-CCSD calculation (eV)} \\
\hline
Valence & & 0.47 & 0.27 & 0.17 & 0.17 & 0.48 & 2.94 & 3.15 & - \\
Rydberg & & 0.99 & 0.07 & 0.06 & 0.06 & 1.45 & 2.03 & 2.04 & - \\
Total   & & 0.84 & 0.13 & 0.09 & 0.09 & 1.17 & 2.29 & 2.36 & - \\
\hline
\multicolumn{10}{c}{Maximum absolute deviation of excitation energies with respect to EOM-CCSD (eV)} \\
\hline
 & & 1.21 & 0.45 & 0.33 & 0.33 & 1.59 & 3.47 & 3.74 & - \\
\hline\hline
  \end{tabular}
    \label{H2COtable_dyn}
\end{table*}
\endgroup

\begingroup
\squeezetable
\begin{table*}
  \centering
  \caption
    {Same as Table~\ref{N2table_dyn} for C$_2$H$_4$.}
  \begin{tabular}{c l c c H c c H c c}
    \hline\hline
State 		& Transition 	& TDKS & TDRSH	& TDRSH+BSE2' & TDRSH+BSE2 & TDHF	& TDHF+BSE2' & TDHF+BSE2 & EOM-CCSD \\
\hline
\multicolumn{10}{c}{ Valence excitation energies (eV)} \\
\hline
$^3B_{1\text{u}}$      & $1b_{3\text{u}}(\pi) \rightarrow 1b_{2\text{g}} (\pi^*)$                    & 4.74  & 4.35  & 4.73  & 4.73  & 3.54  & 5.92  & 6.06  & 4.41\\
$^1B_{1\text{u}}$      & $1b_{3\text{u}}(\pi) \rightarrow 1b_{2\text{g}} (\pi^*)$                    & 7.91  & 8.07  & 8.37  & 8.38  & 7.70  & 9.05  & 9.11  & 8.00\\
$^3B_{1\text{g}}$      & $1b_{3\text{g}}(\sigma) \rightarrow 1b_{2\text{g}}(\pi^*)$                  & 7.18  & 7.92  & 8.04  & 8.04  & 8.48  & 10.33 & 10.43 & 8.21\\
$^1B_{1\text{g}}$      & $1b_{3\text{g}}(\sigma) \rightarrow 1b_{2\text{g}}(\pi^*)$                  & 7.48  & 8.04  & 8.24  & 8.24  & 9.23  & 10.74 & 10.81 & 8.58\\
\hline                                                                                                                                          
\multicolumn{10}{c}{Rydberg excitation energies (eV)} \\                                                                                        
\hline                                                                                                                                          
$^3B_{3\text{u}}$      & $1b_{3\text{u}}(\pi) \rightarrow 4a_{1\text{g}}(\sigma)$                    & 6.59  & 7.21  & 7.35  & 7.35  & 6.91  & 7.36  & 7.37  & 7.16\\
$^1B_{3\text{u}}$      & $1b_{3\text{u}}(\pi) \rightarrow 4a_{1\text{g}}(\sigma)$                    & 6.65  & 7.36  & 7.48  & 7.48  & 7.14  & 7.42  & 7.43  & 7.30\\
$^3B_{1\text{g}}$      & $1b_{3\text{u}}(\pi) \rightarrow 2b_{2\text{u}}(\sigma)$                    & 6.98  & 7.42  & 7.77  & 7.78  & 7.66  & 8.10  & 8.10  & 7.91\\
$^3B_{2\text{g}}$      & $1b_{3\text{u}}(\pi) \rightarrow 3b_{1\text{u}}(\sigma)$                    & 7.10  & 8.03  & 8.11  & 8.11  & 7.79  & 8.06  & 8.07  & 7.93\\
$^1B_{1\text{g}}$      & $1b_{3\text{u}}(\pi) \rightarrow 2b_{2\text{u}}(\sigma)$                    & 7.19  & 7.92  & 8.17  & 8.17  & 7.75  & 8.09  & 8.09  & 7.97\\
$^1B_{2\text{g}}$      & $1b_{3\text{u}}(\pi) \rightarrow 3b_{1\text{u}} (\sigma)$                   & 7.15  & 8.13  & 8.20  & 8.20  & 7.92  & 8.07  & 8.07  & 8.01\\
$^3A_\text{\text{g}}$         & $1b_{3\text{u}}(\pi) \rightarrow 2b_{3\text{u}}(\pi)$                       & 8.03  & 8.46  & 8.60  & 8.60  & 8.02  & 8.62  & 8.64  & 8.48\\
$^1A_\text{\text{g}}$         & $1b_{3\text{u}}(\pi) \rightarrow 2b_{3\text{u}}(\pi)$                       & 8.30  & 8.87  & 8.99  & 8.99  & 8.61  & 8.88  & 8.88  & 8.78\\
$^3B_{3\text{u}}$      & $1b_{3\text{u}}(\pi) \rightarrow 5a_{1\text{g}}(\sigma)$                    & 8.26  & 8.97  & 9.12  & 9.12  & 8.74  & 9.26  & 9.26  & 9.00\\
$^1B_{3\text{u}}$      & $1b_{3\text{u}}(\pi) \rightarrow 5a_{1\text{g}}(\sigma)$                    & 8.28  & 9.09  & 9.20  & 9.20  & 8.92  & 9.13  & 9.13  & 9.07\\
\hline 
\multicolumn{10}{c}{ Ionization threshold: $-\epsilon_{\text{HOMO}}$ (eV)} \\
\hline
 & & 6.89 & 10.45 &       &       & 10.23 &       &       & \\
\hline
\multicolumn{10}{c}{ MAD of excitation energies with respect to the EOM-CCSD calculation (eV)} \\
\hline
Valence & & 0.64 & 0.24 & 0.30 & 0.30 & 0.52 & 1.71 & 1.80 & - \\
Rydberg & & 0.71 & 0.10 & 0.17 & 0.17 & 0.21 & 0.14 & 0.14 & - \\
Total   & & 0.69 & 0.14 & 0.21 & 0.21 & 0.30 & 0.59 & 0.62 & - \\
\hline
\multicolumn{10}{c}{Maximum absolute deviation of excitation energies with respect to EOM-CCSD (eV)} \\
\hline
 & & 1.10 & 0.54 & 0.37 & 0.38 & 0.87 & 2.16 & 2.23 & - \\
\hline\hline
  \end{tabular}
    \label{C2H4table_dyn}
\end{table*}
\endgroup

Figure~\ref{fig:be} shows the long-range excitation energies to the first triplet and singlet excited states of the Be atom. For these low-lying valence states, the TDHF long-range excitation energies are relatively accurate close to $\mu=0$, but they deteriorate somewhat as $\mu$ is increased. For $\mu \gtrsim 1$ bohr$^{-1}$, TDHF underestimates the triplet long-range excitation energy by about 1 eV and the singlet long-range excitation energy by about 0.25 eV, in comparison to the reference FCI long-range excitation energies. Adding the BSE2 correlation kernel correctly increases the TDHF long-range excitation energies, leading to a relatively accurate singlet excitation energy for all $\mu$ and reducing the error in the TDHF triplet excitation energy by a factor of 3 for large $\mu$. The GW2 quasiparticle correction does not change much the excitation energies, at most about 0.1 eV.

Finally, the long-range excitation energies to the first triplet and singlet excited states of the H$_2$ molecule at the equilibrium internuclear distance are reported in Figure~\ref{fig:be}. For these molecular valence states, the LDA orbital energy differences at $\mu=0$ are too low by more than 1 eV. Again, this underestimation is corrected by increasing the value of $\mu$. For $\mu \gtrsim 1$ bohr$^{-1}$, TDHF always underestimates the triplet long-range excitation energy, while it is more accurate for the singlet long-range excitation energy. The addition of the BSE2 correlation kernel changes little the excitation energy for the singlet state, and significantly reduces the error on the excitation energy for the triplet excited state. The GW2 quasiparticle correction tends to improve a bit the accuracy of the excitation energies for intermediate values of $\mu$, but remains very small for small and large values of $\mu$ (in particular, it is about 0.05 eV for $\mu=0.35$ bohr$^{-1}$).

Overall, these results are encouraging and support the relevance of the TDHF+BSE2 approximation for the long-range response kernel, as well as the neglect of the GW2 quasiparticle correction for small enough value of $\mu$.

\begin{figure}
\includegraphics[scale=0.30,angle=-90]{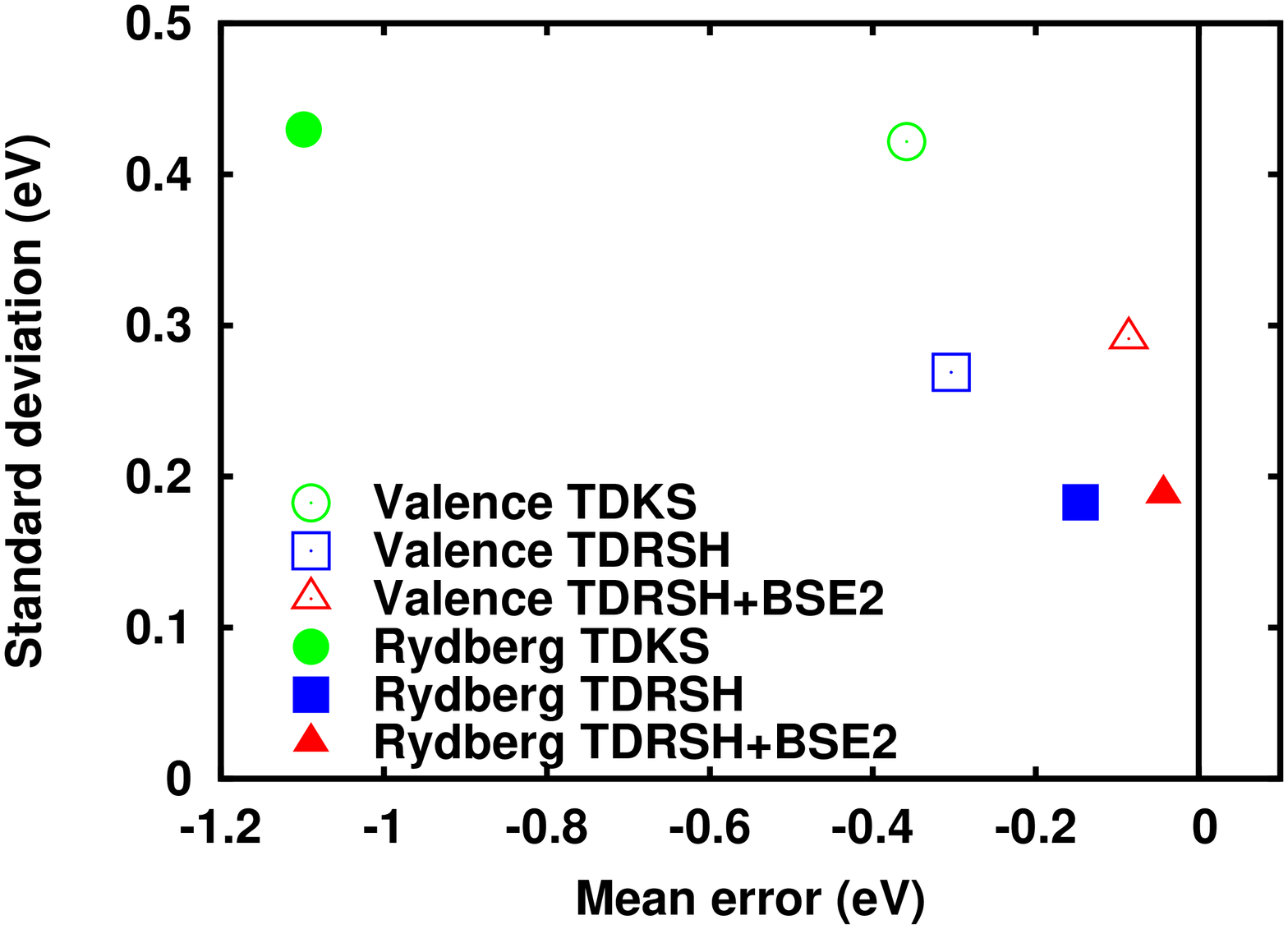}
\caption{Mean error versus standard deviation for the valence and Rydberg excitation energies of the N$_2$, CO$_2$, H$_2$CO, and C$_2$H$_4$ molecules calculated with linear-response TDKS (with the LDA functional), TDRSH and TDRSH+BSE2 (with the short-range LDA functional and $\mu=0.35$ bohr$^{-1}$), all within the TDA. The errors are calculated with respect to the EOM-CCSD excitation energies. The Sadlej+ basis set is used.}
\label{fig:stat}
\end{figure}

\subsection{Excitation energies of the N$_2$, CO$_2$, H$_2$CO, and C$_2$H$_4$ molecules}
\label{sec:results_more}

We now test the calculation of excitation energies with the complete proposed TDHF+BSE2 method, i.e. including now the short-range Hxc kernel in the linear-response part and neglecting the GW2 quasiparticle correction. The excitation energies for each method and each molecule are given in Tables~\ref{N2table_dyn}-\ref{C2H4table_dyn}. Mean absolute deviations (MAD) and maximum absolute deviations with respect to the EOM-CCSD reference are also given for valence, Rydberg, and all excitation energies. 

As already well known, TDKS with the LDA functional gives very underestimated Rydberg excitation energies. TDRSH greatly improves the excitation energies for the Rydberg states and, to a lesser extent, for the valence states, resulting in total MADs of 0.41, 0.26, 0.13, and 0.14 eV for N$_2$, CO$_2$, H$_2$CO, and C$_2$H$_4$, respectively. TDRSH also offers a more accurate description of valence and Rydberg excitation energies than TDHF. For a more intensive discussion of the performance of TDRSH, see Ref.~\onlinecite{RebSavTou-MP-13}.

Both when starting from TDHF and TDRSH, the addition of the BSE2 correlation correction always leads to larger excitation energies. In the case of TDHF, the BSE2 correction increases the valence excitation energies by about 2 or 3 eV, leading to largely overestimated valence excitation energies. For Rydberg states, the BSE2 correction on top of TDHF is smaller (less than 1 eV) but also leads to systematically overestimated excitation energies. Overall, TDHF+BSE2 considerably worsens the TDHF excitation energies. We thus conclude that, for these molecules, the relatively accurate results reported by Zhang \emph{et al.}~\cite{ZhaSteYan-JCP-13} crucially depend on using the GW2 quasiparticle correction to the HF orbital energies.

In the range-separated case, the long-range BSE2 correction induces only a moderate increase of the valence excitation energies by about 0.2 to 0.4 eV, leading for these states to MADs of 0.35, 0.16, 0.17, and 0.30 eV for N$_2$, CO$_2$, H$_2$CO, and C$_2$H$_4$, respectively. The TDRSH+BSE2 excitation energies of the Rydberg states are also systematically larger than the TDRSH ones by usually less than 0.1 eV, giving for these states MADs of 0.27, 0.22, 0.06, and 0.17 eV for N$_2$, CO$_2$, H$_2$CO, and C$_2$H$_4$, respectively. Of course, the difference in magnitude of the BSE2 correction in the range-separated case compared to the full-range case is to be mostly attributed to the substitution of the full-range two-electron integrals by the long-range ones. Since for the chosen value of the range-separation parameter $\mu$ of 0.35 bohr$^{-1}$, TDRSH mostly gives slightly underestimated excitation energies of the considered systems, the long-range BSE2 correction overall slightly improves the excitation energies. More specifically, in comparison with TDRSH, TDRSH+BSE2 gives slightly smaller total MADs of 0.32, 0.19, and 0.09 eV for N$_2$, CO$_2$, and H$_2$CO, and a slightly larger MAD of 0.21 eV for C$_2$H$_4$. Also, for all the four molecules, TDRSH+BSE2 always gives the smallest maximum absolution deviation among all the methods, suggesting that TDRSH+BSE2 describes more reliably the excitation energies than the other methods.

Finally, as a global summary of the results, Figure~\ref{fig:stat} reports the mean error versus the standard deviation for the valence and Rydberg excitation energies of the four molecules for the different methods. For the valence excitation energies, going from TDKS to TDRSH mainly decreases the standard deviation, while going from TDRSH to TDRSH+BSE2 decreases the mean error. For the Rydberg excitation energies, TDRSH provides a large improvement over TDKS both in terms of mean error and standard deviation, while TDRSH+BSE2 gives a slightly smaller mean error than TDRSH.

\section{Conclusion}
\label{sec:conclu}

We have developed a range-separated linear-response TDDFT approach using a long-range frequency-dependent second-order Bethe-Salpeter correlation kernel. We have tested our approach using a perturbative resolution of the linear-response equations within the TDA for valence and Rydberg excitation energies of small atoms and molecules. The results show that the addition of the long-range correlation kernel overall slightly improves the excitation energies. 

More intensive tests should now be carried out with this long-range correlation kernel to better assess its performance. In particular, this long-range correlation kernel is expected to be appropriate for (1) calculating excitation energies of excited states with significant double-excitation contributions, (2) calculating charge-transfer excitation energies, and (3) calculating dispersion interactions in excited states.

A number of further developments should also be explored: adding the self-energy (or quasiparticle) contribution directly to the kernel, going beyond the TDA and the perturbative resolution of the linear-response equations, and going beyond the second-order approximation. Finally, we note that the present work could be repeated using a linear decomposition of electron-electron interaction~\cite{ShaTouSav-JCP-11}, instead of a range separation, in order to construct a new TDDFT double-hybrid method which would be an alternative to the one commonly used based on CIS(D)~\cite{GriNee-JCP-07}.

\section*{Acknowledgements}
We thank J. A. Berger, E. Luppi, D. Mukherjee, L. Reining, P. Romaniello, and A. Savin for discussions.

\appendix
\section{Non-interacting four-point linear response function}
\label{app}

In this appendix, we derive the expression of the non-interacting linear-response function $\chi_0(\omega',\omega)$ depending on two frequencies which is used in Eq.~(\ref{chi0iaia2omega}).

The non-interacting four-point linear-response function is defined in the time domain by
\begin{eqnarray}
\chi_0(\mx_1t_1,\mx_2t_2; \mx_1't_1',\mx_2't_2') = \;\;\;\;\;\;\;\;\;\;\;\;\;\;\;\;\;\;\;\;\;\;\;\;\;\;\;
\nonumber\\
\;\;\;\;\;\;\;\;\;\;\;\;\;\;\;\;\;\;\;\;\;\;\;\;\;\;\; -i \, G_0 (\mx_1t_1,\mx_2't_2') G_0 (\mx_2t_2,\mx_1't_1'),
\label{chi0GG}
\end{eqnarray}
where $G_0$ is the non-interacting one-electron Green function. Using time-translation invariance and introducing the time variables $\tau_1=t_1-t_1'$, $\tau_2=t_2-t_2'$, and $\tau=(t_1 +t_1')/2-(t_2 +t_2')/2$, Eq.~(\ref{chi0GG}) becomes
\begin{eqnarray}
\chi_0(\mx_1,\mx_2; \mx_1',\mx_2';\tau_1,\tau_2,\tau) = \;\;\;\;\;\;\;\;\;\;\;\;\;\;\;\;\;\;\;\;\;\;\;\;\;\;\;\;\;\;\;\;
\nonumber\\
 -i \, G_0 \left(\mx_1,\mx_2',\tau + \frac{\tau_1+\tau_2}{2}\left) G_0 \right(\mx_2,\mx_1',\frac{\tau_1+\tau_2}{2}-\tau\right),
\nonumber\\
\end{eqnarray}
with $G_0 (\mx_1,\mx_1',t_1-t_1')=G_0 (\mx_1t_1,\mx_1't_1')$. The triple Fourier transform of $\chi_0(\mx_1,\mx_2; \mx_1',\mx_2';\tau_1,\tau_2,\tau)$ is easily found to be
\begin{eqnarray}
\chi_0(\mx_1,\mx_2; \mx_1',\mx_2';\omega',\omega'',\omega) = 
\nonumber\\
\int \d\tau_1 \d\tau_2 \d\tau \, e^{i\omega' \tau_1} e^{i\omega'' \tau_2} e^{i\omega \tau} \chi_0(\mx_1,\mx_2; \mx_1',\mx_2';\tau_1,\tau_2,\tau) =
\nonumber\\
-2\pi i \delta(\omega'-\omega'') \,  G_0 \left(\mx_1',\mx_2',\omega'+\frac{\omega}{2}\right) G_0 \left(\mx_2,\mx_1',\omega'-\frac{\omega}{2}\right),
\nonumber\\
\label{ch0omegapomegappomega}
\end{eqnarray}
where $G_0 (\mx_1,\mx_1',\omega) = \int \d\tau_1 e^{i\omega \tau_1} G_0 (\mx_1,\mx_1',\tau_1)$ is the Fourier transform of the Green function. The linear-response function $\chi_0(\mx_1,\mx_2; \mx_1',\mx_2';\omega',\omega)$ depending on two frequencies is then given by
\begin{eqnarray}
\chi_0(\mx_1,\mx_2; \mx_1',\mx_2';\omega',\omega)
\nonumber\\
=\chi_0(\mx_1,\mx_2; \mx_1',\mx_2';\tau_1=0^-,\omega',\omega) 
\nonumber\\
=\chi_0(\mx_1,\mx_2; \mx_1',\mx_2';\omega',\tau_2=0^-,\omega) 
\nonumber\\
=\int \frac{\d\omega''}{2\pi} e^{i\omega'' 0^+} \chi_0(\mx_1,\mx_2; \mx_1',\mx_2';\omega',\omega'',\omega)
\nonumber\\
= -i \,  e^{i\omega' 0^+} G_0 \left(\mx_1',\mx_2',\omega'+\frac{\omega}{2}\right) G_0 \left(\mx_2,\mx_1',\omega'-\frac{\omega}{2}\right).
\nonumber\\
\label{ch0omegapomega}
\end{eqnarray}
Inserting in Eq.~(\ref{ch0omegapomega}) the Lehmann representation of the Green function,
\begin{eqnarray}
G_0(\mx_1,\mx_1',\omega) = \sum_a \frac{\varphi_a(\mx_1)\varphi_a^*(\mx_1')}{\omega - \varepsilon_a + i 0^+} + \sum_i \frac{\varphi_i(\mx_1)\varphi_i^*(\mx_1')}{\omega - \varepsilon_i - i 0^+},
\nonumber\\
\end{eqnarray}
where $i$ and $a$ refer to occupied and virtual spin orbitals, respectively, and using the identity $1/[(\omega'-A)(\omega'-B)]=[1/(\omega'-A) - 1/(\omega'-B)]/(A-B)$ gives
\begin{eqnarray}
\chi_0(\mx_1,\mx_2; \mx_1',\mx_2';\omega',\omega) = \;\;\;\;\;\; \;\;\;\;\;\;\;\;\;\;\;\; \;\;\;\;\;\;
\nonumber\\
i \,  e^{i\omega' 0^+} \Biggl[ \sum_{ia} \dfrac{\varphi_i^*(\mx_1')\varphi_a(\mx_1)\varphi_a^*(\mx_2')\varphi_i(\mx_2)}{\omega -(\varepsilon_a - \varepsilon_i) +i0^+} 
\nonumber\\
\times \left( \frac{1}{\omega'+\omega/2-\varepsilon_a +i0^+} - \frac{1}{\omega'-\omega/2-\varepsilon_i -i0^+}\right)
\nonumber\\
+\sum_{ia} \dfrac{\varphi_a^*(\mx_1')\varphi_i(\mx_1)\varphi_i^*(\mx_2')\varphi_a(\mx_2)}{-\omega -(\varepsilon_a - \varepsilon_i) +i0^+}
\nonumber\\
\times \left( \frac{1}{\omega'-\omega/2-\varepsilon_a +i0^+} - \frac{1}{\omega'+\omega/2-\varepsilon_i -i0^+}\right)
\nonumber\\
+\sum_{ab} \dfrac{\varphi_a^*(\mx_1')\varphi_b(\mx_1)\varphi_b^*(\mx_2')\varphi_a(\mx_2)}{\omega -(\varepsilon_b - \varepsilon_a)} 
\nonumber\\
\times \left( \frac{1}{\omega'+\omega/2-\varepsilon_b +i0^+} - \frac{1}{\omega'-\omega/2-\varepsilon_a +i0^+}\right)
\nonumber\\
+\sum_{ij} \dfrac{\varphi_i^*(\mx_1')\varphi_j(\mx_1)\varphi_j^*(\mx_2')\varphi_i(\mx_2)}{\omega -(\varepsilon_j - \varepsilon_i)} 
\nonumber\\
\times \left( \frac{1}{\omega'+\omega/2-\varepsilon_j -i0^+} - \frac{1}{\omega'-\omega/2-\varepsilon_i -i0^+}\right)
\Biggl].
\nonumber\\
\label{chi0omegapomegalong}
\end{eqnarray}
In Eq.~(\ref{chi0omegapomegalong}), the first sum corresponds to the matrix element $\chi_{0,ia,ia}(\omega',\omega)$ written in Eq.~(\ref{chi0iaia2omega}), while the second sum corresponds to the matrix element $\chi_{0,ai,ai}(\omega',\omega) = \chi_{0,ia,ia}(\omega',-\omega)$.


\begin{thebibliography}{88}
\bibitem[{Reb()}]{RebTou-JJJ-XX-note1}
\bibinfo{note}{The last two terms of the kernel in Eq. (31) of
  Ref.~\onlinecite{ZhaSteYan-JCP-13} contain non-antisymmetrized two-electron
  integrals. However, these terms can also be written with a factor of $1/2$
  and antisymmetrized two-electron integrals, leading to our Eq.~(\ref{eq:eff
  kernel iajb}).}
\end{thebibliography}
\end{document}